\documentclass[usenatbib]{mn2e}
\usepackage{amsmath}
\usepackage{txfonts}
\usepackage{amsbsy}
\usepackage{psfig}
\usepackage[dvips]{graphicx,color}
\usepackage{natbib}
\usepackage{longtable,lscape} 
\usepackage{microtype}
\bibpunct{(}{)}{;}{a}{}{,}
\DeclareMathAlphabet{\mathsc}{OT1}{cmr}{m}{sc}
\def\testbx{bx}%
\DeclareRobustCommand{\ion}[2]{%
\relax\ifmmode
\ifx\testbx\f@series
{\mathbf{#1\,\mathsc{#2}}}\else
{\mathrm{#1\,\mathsc{#2}}}\fi
\else\textup{#1\,{\mdseries\textsc{#2}}}%
\fi}

\newcommand{\Ha}{H$\alpha$}

\newcommand{\kms}{km~s\ensuremath{^{-1}}}

\def\kms{km\,s$^{-1}$}

\newcommand{\gapp}{\mbox{\raisebox{-0.3em}{$\stackrel{\textstyle >}{\sim}$}}}

\def\zem{$z_{\rm em}$~}

\def\hi{H\,{\sc i}}

\def\kms{km~s$^{-1}$}
\def\ugc{UGC\,07904}

\title[HI gas at pc to kpc scales]
{Revealing \hi\ gas in emission and absorption on pc to kpc scales in a galaxy at $z$$\sim$0.017}
\author[Gupta et.al]{N. Gupta$^{1}$\thanks{E-mail:ngupta@iucaa.in}, 
R. Srianand$^1$, 
J. S. Farnes$^{2,3}$, 
Y. Pidopryhora$^4$, 
M. Vivek$^5$,
Z. Paragi$^6$,
\newauthor P. Noterdaeme$^7$, 
T. Oosterloo$^{8,9}$ and
P. Petitjean$^7$
\\ 
$^{1}$ Inter-University Centre for Astronomy and Astrophysics, Post Bag 4, Ganeshkhind, Pune 411007, India\\
$^{2}$ Department of Astrophysics/IMAPP, Radboud University, PO Box 9010, NL-6500 GL Nijmegen, the Netherlands\\
$^{3}$ Oxford e-Research Centre (OeRC), Keble Road, Oxford OX1 3QG, England  \\
$^{4}$ Max-Planck-Institut f{\"u}r Radioastronomie,  Auf dem H{\"ü}gel 69,  D-53121 Bonn, Germany \\ 
$^{5}$ Department of Physics and Astronomy, University of Utah, Salt Lake City, UT 84112, USA \\ 
$^{6}$ Joint Institute for VLBI ERIC, Postbus 2, 7990 AA Dwingeloo, The Netherlands \\
$^{7}$ UPMC-CNRS, UMR7095, Institut d'Astrophysique de Paris, F-75014 Paris, France \\
$^{8}$ ASTRON, the Netherlands Institute for Radio Astronomy, Postbus 2, 7990 AA Dwingeloo, The Netherlands \\
$^{9}$ Kapteyn Astronomical Institute, University of Groningen, Postbus 800, 9700 AV Groningen, The Netherlands \\
}

\begin{document}

\pagerange{\pageref{firstpage}--\pageref{lastpage}} \pubyear{2017}

\maketitle

\label{firstpage}

\begin{abstract}
We present a detailed study of the quasar-galaxy pair: J1243+4043 - UGC\,07904. The sight line of the background quasar 
( $z_q$= 1.5266) passes through a region of the galaxy ($z_g$=0.0169) at an impact parameter of 6.9\,kpc with high 
metallicity (0.5\,Z$_\odot$) and negligible dust extinction. 
We detect \hi\ 21-cm absorption from the foreground galaxy at arcsecond and milliarcsecond scales.  
For typical cold neutral medium (CNM) temperatures in the Milky Way, this 21-cm absorber can be classified 
as a damped Ly$\alpha$ absorber (DLA). We infer the harmonic mean spin temperature of the gas to be $\sim$400\,K 
and for a simple two-phase medium we estimate the CNM-fraction to be $f_{\rm CNM}$ = 0.27.  This is remarkably consistent 
with the CNM fraction observed in the Galaxy and less than that of high-redshift DLAs.  
The quasar exhibits a core-jet morphology on milliarcsecond scales, 
corresponding to an overall extent of $\sim$9\,pc at $z_g$.
We show that the size of CNM absorbing clouds associated with the foreground galaxy is $>$5\,pc and 
they may be part of cold gas structures that extend beyond $\sim$35\,pc.     
Interestingly, the rotation measure of quasar J1243+4043 is higher than any other source in samples of quasars 
with high-$z$ DLAs. 
However, we do not find any detectable differences in RMs and polarization fraction of sight lines with or 
without high-$z$ ($z\ge2$) DLAs  or low-$z$ ($z\le0.3$) 21-cm absorbers.  
Finally, the foreground galaxy UGC\,07904 is also part of a galaxy group.  We serendipitously detect \hi\ 21-cm emission from four 
members of the group, and a $\sim$80\,kpc long \hi\ bridge connecting two of the other members. 
The latter, together with the properties of the group members, suggests that the group is a highly interactive 
environment. 

\end{abstract}


\date{Received; accepted}

\section{Introduction}

At cosmologically significant redshifts, the bulk of \hi\ in galaxies is probed via Damped Ly$\alpha$ 
Absorbers (DLAs), defined to have $N$(\hi)$\ge$2$\times$$10^{20}$\,cm$^{-2}$, seen in the optical spectra 
of distant quasars \citep[][]{Noterdaeme12dla}.  
Due to the atmospheric cut-off in ultraviolet, large samples of DLAs are only available at $z\gapp$1.6. 
Compared with the $\Omega_{\rm {HI}}$($z = $0) measured from 21-cm emission line observations, its decrease from 
$z$$\sim$2 is less than a factor of 2 which is very modest compared to the order of magnitude decrease in 
the star-formation rate (SFR) density over the same redshift range \citep[][]{Madau14, Hoppmann15}. This implies that the processes 
leading to the conversion of atomic gas into molecular gas and eventually into stars need to be understood 
directly via observations of cold atomic and molecular gas \citep[e.g.][]{Curran17}, rather than via 
the evolution of total atomic gas content.

Observations of the 21-cm absorption line which is an excellent tracer of cold atomic gas ($\sim$100\,K) in 
the interstellar medium (ISM) can be used to map the evolution of cold gas in galaxies.  There have been a 
number of \hi\ 21-cm 
absorption line surveys in the past but due to technical limitations most of these have been based on samples 
of high-$z$ Mg~{\sc ii} absorbers and DLAs \citep[e.g.][]{Briggs83, Carilli96, Gupta09, Kanekar09mg2, 
Curran10, Srianand12dla, Gupta12, Kanekar14, Dutta17mg2}.  At these redshifts, it is challenging to determine the 
properties of absorbing galaxies and, hence, the origin of absorbing gas.  Furthermore, due to the 
sparse availability of suitable low-frequency receivers at Very Long Baseline Interferometry (VLBI) antennas one 
cannot carry out milliarcsecond (mas) - scale interferometry to determine the parsec-scale structure of the 
absorbing clouds. The structure and the size distribution of neutral gas are required for determining the true 
21-cm absorption optical depth and spin temperature \citep[][]{Briggs83}, and relevant for understanding 
the processes that determine the stability of these clouds \citep[e.g.][]{Maclow04}.

The subarcsecond-scale spectroscopy of high-$z$ 21-cm absorbers will be possible only after the Square Kilometer 
Array (SKA)- mid and, eventually, the SKA-VLBI with suitable low frequency ($<$1\,GHz) receivers are built 
\citep[][]{Paragi15}.
For the moment, the above-mentioned difficulties can be overcome at low-$z$ by targeting quasar-galaxy 
pairs\footnote{Defined as the fortuitous alignment of a foreground galaxy with a distant background quasar.} 
(QGPs) where the foreground galaxy is at a redshift (typically $z$$<$0.2) such that the redshifted \hi\ 21-cm 
line is observable using VLBI.  
The 21-cm absorption observations of such QGPs covering a wide range of galaxy types and 
environments can be used to build a sample of DLAs at low-$z$ where a direct connection between 
the galaxies and absorbers can be made \citep[e.g.][]{Haschick83, Carilli92, Boisse98, Borthakur10, Gupta10, Gupta13, 
Reeves15, Reeves16, Borthakur16, Dutta17}. 
Furthermore, if the background quasar has structures on parsec scales, the VLBI spectroscopy can be used to probe the 
parsec scale structures in the cold atomic gas \citep[e.g.][]{Srianand13dib, Biggs16}.

\begin{figure}
\hbox{
\includegraphics[width=85mm, angle=0]{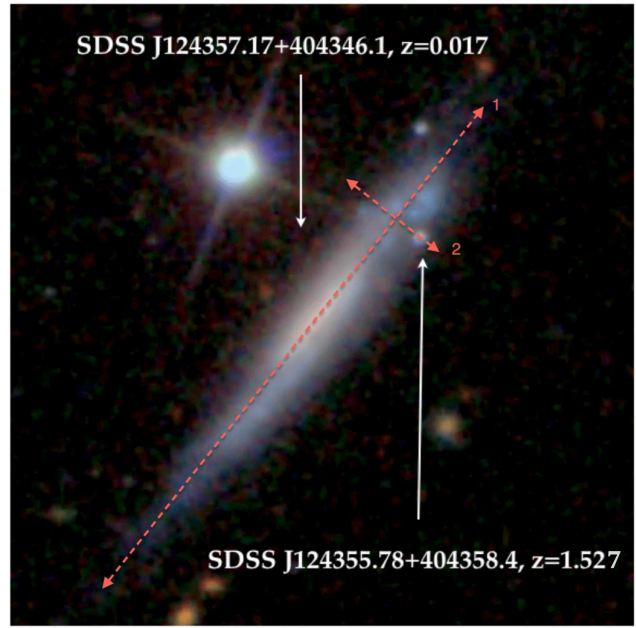}
}
\caption{SDSS color representation (100$^{\prime\prime}$$\times$100$^{\prime\prime}$) 
of the quasar (SDSS J124355.78+404358.4, $z_q$= 1.527) - 
galaxy (SDSS J124357.17+404346.1/\ugc, $z_g$=0.017) pair.  The quasar sight line passes through the galaxy 
at an impact parameter of 20.3$^{\prime\prime}$ (6.9\,kpc at the $z_g$) as measured from the center 
of the galaxy.  Dashed lines mark the orientation of the slits used in the IGO observations. 
}
\label{overlay}
\end{figure}

In this paper, we present a detailed \hi\ 21-cm emission and absorption line analysis of a low-$z$ 
quasar/galaxy pair, SDSS J124357.5+404346.5/ SDSS J124355.78+404358.5, that has above-mentioned  properties.   
Hereafter we refer to this pair as QGP J1243+4043.  
The \hi\ 21-cm absorption towards this QGP was reported in \citet[][]{Dutta17}. 
The quasar J1243+4043 is unique in the sense that it is polarized and, therefore, also offers an 
unique opportunity to probe the magnetoionic plasma from the foreground 21-cm absorbing galaxy along the 
same line of sight.  
Fig.~\ref{overlay} shows the SDSS colour composite image of this QGP.  
The quasar is compact (deconvolved size $<$2$^{\prime\prime}$) 
in the Faint Images of the Radio Sky at Twenty-Centimeters (FIRST) survey and has a flux density of 194\,mJy.  
While the galaxy is well known in literature as UGC\,07904 and has a redshift of $z_g$=0.017 \citep[][]{Nilson73}, 
the redshift of the quasar is unknown from previous literature. 
The quasar sight line passes through the edge of the spiral arm of this nearly edge-on 
galaxy at an impact parameter of 6.9\,kpc\footnote{Throughout 
this paper we use the $\Lambda$CDM cosmology with $\Omega_m$=0.27, $\Omega_\Lambda$=0.73 and 
H$_{\rm o}$=71\,\kms\,Mpc$^{-1}$.} (Fig.~\ref{overlay}).  
\citet{White99} identify the foreground galaxy as a member of a poor galaxy cluster.  
The other three members (not seen in Fig.~\ref{overlay}) of this group are 
UGC\,07921/IC\,3726 ($z$=0.0168),  
IC\,3723 ($z$=0.0179), 
and SDSS J124423.25+404148.5 ($z$=0.0180).  
All the redshifts are from the SDSS database, and the mean redshift of the galaxy group is 0.0174.  

To study this system in detail, 
we have performed long-slit spectroscopic observations of the QGP J1243+4043 using 2-m 
optical telescope at IUCAA\footnote{Inter-University Centre for Astronomy And Astrophysics} Girawali Observatory (IGO) 
to measure the quasar redshift, line of sight reddening and the properties of the ionized gas in the galaxy. 
We have used radio data from the Giant Metrewave Radio Telescope (GMRT), the Westerbork Synthesis 
Radio Telescope (WSRT), and the global-VLBI array consisting of the  Very Long Baseline Array (VLBA) and 
the European VLBI Network (EVN) to map the 
large scale \hi\ 21-cm emission from the QGP and the associated galaxy group, and detect 21-cm 
absorption towards the background quasar at arcsecond and mas scales.  


The layout of this paper is as follows.  
In Section~\ref{sec:data}, we present details of optical and radio data used for this study.  
Results and discussion are presented in Section~\ref{sec:disc}.
We also discuss the unusually high RM of quasar J1243+4043 in the context of other polarized 
quasars with DLAs and 21-cm absorbers.
A summary of the results is presented in Section~\ref{sec:summ}.

\section[]{Observations and data reduction}
\label{sec:data}

\subsection{IGO optical long-slit spectroscopy}

%
\begin{table}
\centering
\caption{Details of IGO observations }
\begin{tabular}{ccccccc}
\hline
{\large \strut} Slit Orientation & Date &  Exposure time  &  Air mass & Grism \\
\hline           
{\large \strut}    1        & 2010/4/11 & 45 &1.169 & GR7  \\ 
                  ''        & 2010/4/11 & 45 &1.100 & GR7  \\
                  ''        & 2010/4/11 & 45 &1.107 & GR7  \\
                   2        & 2010/4/12 & 45 &1.500 & GR7  \\
                  ''        & 2010/4/14 & 30 &1.190 & GR7  \\
                  ''        & 2010/4/14 & 45 &1.137 & GR7  \\
                  ''        & 2010/4/14 & 45 &1.086 & GR8  \\
                  ''        & 2010/4/14 & 45 &1.077 & GR8  \\
\hline
\end{tabular}
\flushleft{Column 1: slit orientation as shown in Fig.~\ref{overlay}; column 2: date of observations in format yyyy/mm/dd; column 3: exposure time in min; 
column 4 and 5: air mass and grism respectively.}
\label{logobs}
\end{table}

The IGO observations were performed on 2010, April 11, 12 and 14 using 
the IUCAA Faint Object Spectrograph (IFOSC) with slit orientations as shown 
in Fig.~\ref{overlay}.  The observing details are provided in Table~\ref{logobs}. 
The seeing, measured from the images taken at the night, was 1.2$\arcsec$-1.6\,$\arcsec$.   
The length of the long slit is 10.5$^{\prime}$.
The slit width was kept at 1.5$\arcsec$.  
In orientation 1, the slit was aligned along the disk of the galaxy, 
whereas in orientation 2, it was aligned to cover the quasar and trace the properties of 
gas perpendicular to the galaxy disk.
In each slit orientation, we used grisms GR7 and GR8 of the IFOSC  
to obtain the spectra over the wavelength range of 3800-9000\,\AA.   

The data were reduced using a Helium-Neon lamp spectrum as the comparison 
and by following standard procedures using {\tt IRAF} 
involving corrections for geometrical distortions, vacuum wavelength 
calibration, and flux calibration. 
For flux calibration, we used observations of the spectrophotometric standard 
star Feige 34.  Simple flat-fielding did not remove the fringing in 
the red part of the GR8 spectra. To remove this effect for the quasar we 
obtained spectra for slit position 2 by placing the quasar at different 
locations along the slit and subtracting one exposure from the
other \citep[as done in][]{Vivek09}.
Finally, spectra in the heliocentric frame were extracted at different positions 
along slit orientations 1 and 2 using sub-apertures of dimensions 3.5$\arcsec$$\times$1.5$\arcsec$ and 
1.8$\arcsec$$\times$1.5$\arcsec$ respectively. 

\begin{figure*}
\includegraphics[width=180mm,height=90mm]{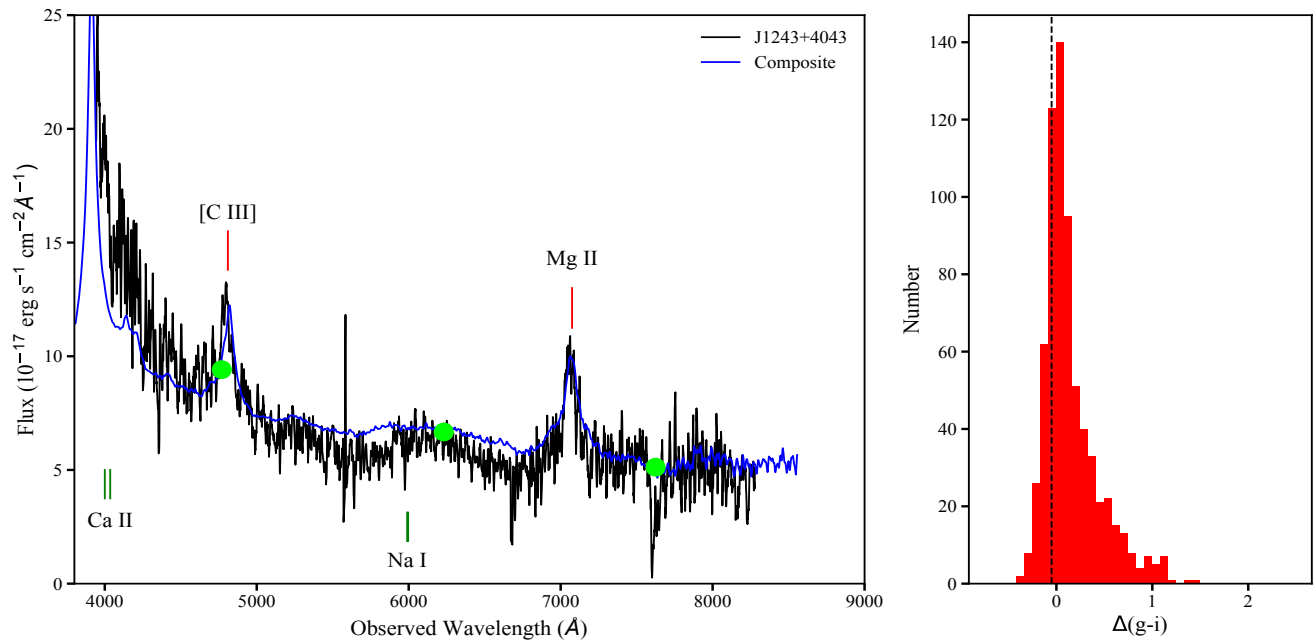}
\caption{{\it Left:} The  IUCAA Girawali Observatory  spectrum of the quasar J1243+4043. 
Filled circles are broadband fluxes of the quasar from the SDSS.
The SDSS composite quasar spectrum from \citet{VandenBerk01}, as redshifted to \zem = 1.5266, is overlaid.
The expected positions of Na~{\sc i} and Ca~{\sc ii} absorption lines at the redshift
of \ugc\ are also marked. 
{\it Right:} The distribution of $\Delta$(g-i) for 3000 quasars at \zem$\sim$1.52 from the SDSS.
The $\Delta$(g-i) for the quasar J1243+4043 is -0.05$\pm$0.02. 
}
\label{qspec}
\end{figure*}

\subsection{GMRT observations}
\label{sec:gmrt}
%
We observed the QGP J1243+4043 using the GMRT 1.4\,GHz receiver on 2010, July, 1. 
The pointing center was the position of the quasar.  The baseband was centered 
at the redshifted 21-cm absorption frequency corresponding to the galaxy 
redshift, $z_g$ = 0.0169, as measured from the SDSS spectrum. 
We observed the standard flux density calibrator 3C\,286 for 10 to 15\,min 
every 2-3 hr to obtain a reliable flux and bandpass calibration.  The compact 
radio source J1227+365 was also observed for 7\,min every $\sim$50\,min for 
phase calibration.  The total on-source time was $\sim$4.6\,hrs. 
The data were acquired using both the hardware and software backends. 

The \hi\ 21-cm absorption spectrum obtained using the data from GMRT hardware correlator 
are presented in \citet[][]{Dutta17}.  Here we present the higher spectral resolution 
data with a $\sim$4\,MHz bandwidth split into 512 channels, acquired through the 
GMRT software backend (GSB).
These data were reduced using the Automated Radio Telescope Imaging Pipeline ({\tt ARTIP}) that is being 
developed to perform the end-to-end processing (i.e. from the ingestion of the raw visibility data 
to the spectral line imaging) 
of data from the uGMRT and MeerKAT absorption line surveys.  The pipeline is written 
using standard python libraries and the {\tt CASA}
package.  Details will be presented in a future paper.  
In short, following data ingestion the pipeline automatically identifies bad antennas, 
baselines, time ranges and radio frequency interference (RFI), using directional and median absolute deviation (MAD) statistics.  
After excluding these bad data, the complex antenna gains as a function of time and 
frequency are determined using the standard flux/bandpass and phase calibrators.  
Applying these gains, a continuum map that uses a user-defined range of frequency channels 
(200-450 - excluding channels with absorption, in the case of QGP J1243+4043) is made. Using this map as a model, self-calibration 
complex gains are determined and then applied to all the frequency channels. 
Finally, the self-calibrated continuum map of J1243+4043 was made using {\tt CASA} task {\tt tclean} 
with {\tt ROBUST=0.5} visibility weighting. The map has a synthesized beam of 
2.5$^{\prime\prime}$$\times$2.1$^{\prime\prime}$ and an rms of 0.4\,mJy\,beam$^{-1}$. 
The quasar is compact at this resolution and has a peak flux density of $\sim$206\,mJy\,beam$^{-1}$.
The pipeline used the CLEAN component model based on this map to subtract the continuum emission from the 
{\it uv} data. This continuum-subtracted data set was then imaged to obtain the spectral-line 
cube.  The stokes $I$ spectrum extracted from this cube at the location of quasar J1243+4043  
has a resolution of 1.7\,\kms\ and an rms of 1.8\,mJy\,beam$^{-1}$\,channel$^{-1}$.

\subsection{WSRT observations}
\label{sec:wsrt}
%
The QGP J1243+4043 was observed with the WSRT at 1.4\,GHz in Maxi-Short configuration 
using a baseband bandwidth of 10\,MHz split into 2048 frequency channels. 
The telescope was pointed at the quasar coordinates.   
The total on-source time, split over five observing runs between 2010, October, 25 
and 2011, January, 29, is $\sim$21\,hr.  
The standard flux density calibrators  3C\,147 and 3C\,286 were observed at the 
beginning and the end of each observing run for flux and bandpass calibration.  
The editing of bad data, and the flux and bandpass calibration were done using the 
{\tt AIPS}. 
After this, the {\it uv} data from different observing runs were concatenated using 
the {\tt AIPS} task {\tt DBCON}.  
An initial continuum image was made using line-free channels, and this was then self-calibrated.  
The image was made using {\tt ROBUST=0} weighting, and has a beam of 
30.7$^{\prime\prime}$$\times$13.9$^{\prime\prime}$ 
with a position angle = 31$^{\circ}$.  
The quasar J1243+4043 in this map has a flux density of $\sim$200\,mJy, consistent with the GMRT observations. 
The complex gains from self-calibration were applied to the line {\it uv} dataset. 
The radio continuum-subtracted line dataset was then shifted to the heliocentric frame 
using the {\tt AIPS} task {\tt CVEL} and also Hanning smoothed to a velocity resolution 
of $\sim$4\,\kms\ (i.e. similar to the resolution of the GMRT dataset used for the analysis).  

We imaged this line dataset using the {\tt CASA}
package to obtain \hi\ cubes and total \hi\ maps.  
For the 21-cm absorption line analysis, we made an \hi\ cube using the same weighting and beam size 
as the continuum image. 
The \hi\ cube was {\tt CLEAN}ed down to twice the single channel noise using image masks.  
The masks were iteratively enlarged after each major cycle of {\tt CLEAN} to include the 
pixels with detectable extended emission as determined from the \hi\ cubes spatially smoothed 
using a Gaussian kernel. 
The cube has a single channel rms of 0.6\,mJy\,beam$^{-1}$. 
Due to the excellent low-surface brightness sensitivity of the WSRT, 
in the {\tt ROBUST=0} \hi\ cube, we detect both 21-cm emission and absorption 
from \ugc\ towards the quasar J1243+4043.  
In addition, we also detect \hi\ emission from the other members of the galaxy group (see Section~\ref{sec:kin}).    

The total intensity \hi\ map was made by summing over the emission detected in the \hi\ cube.   
To exclude noise pixels from the summation and detect diffuse/faint emission at the 
edges, the summation was done as follows. 
First, we created a mask with values of 0 or 1 based on \hi\ emission detected at the 4$\sigma$ 
level either in 
(1) the original \hi\ cube or (2) the cube smoothed using a Gaussian filter of FWHM 
60$^{\prime\prime}$$\times$30$^{\prime\prime}$ with a position angle of 31$^\circ$ and a Hanning 
filter of width $\sim$8\,\kms. This mask was then convolved with the synthesized beam 
of the \hi\ cube and only pixels with values greater than 0.5 in the convolved mask 
were retained. This mask was used to create the total \hi\ intensity map which was  
finally corrected for the effects of the WSRT primary beam attenuation.   

\subsection{Global VLBI 21-cm absorption}
\label{sec:vlbi}

The global VLBI observations of quasar J1243+4043 were carried out on 2011, June, 8.  In total 16 stations, 10 from  
the VLBA and 6 (i.e. Effelsberg, WSRT as a phased array, 
the Jodrell Bank (Lovell) telescope, Onsala-25m, Medicina and Torun) from the EVN were used.  
The interferometric data from the WSRT dishes were also recorded in order to 
measure the total continuum flux density of the quasar at arcsecond scales. The source 3C286 was 
observed for the flux density and bandpass calibrations.  

For VLBI, we observed 4C\,39.25 and 3C\,345 as fringe-finders and 
bandpass calibrators, and J1242+3751 as the phase calibrator.   
The total observing time was 12\,hrs. The EVN participated only for the first 10\,hrs.
The aggregate bit rate per telescope was 128 Mbit/s. We employed 2-bit
sampling and split the data to 4$\times$4\,MHz sub-bands in each of the polarizations
(R, L). The VLBI data were correlated at EVN MkIV Data Processor at 
the Joint Institute for VLBI in Europe (JIVE) in Dwingeloo, the Netherlands in two separate passes with
an averaging time of 2 seconds. The low spectral-resolution {\it continuum pass}
had 16 spectral channels in each sub-band. 
In the {\it line pass} the second band which 
was centered at the redshifted 21-cm frequency was correlated to yield 1024 spectral channels.
The shortest baseline in the data is $\sim$250\,k$\lambda$, where $\lambda$ is the observing wavelength.  
Conservatively, this implies that the data are only sensitive to structures $<1^{\prime\prime}$.
We followed the standard data reduction procedures in {\tt AIPS} to calibrate these data.  
The continuum and line images were made following the steps as outlined in the 
Sections~\ref{sec:gmrt} and \ref{sec:wsrt}.  

The VLBI radio continuum image made using {\tt ROBUST=0} weighting has a beam size and rms 
of 7.0\,mas$\times$2.7\,mas (Position angle=$-$15$^\circ$) 
and $\sim$100\,$\muup$Jy\,beam$^{-1}$ respectively. 

For spectral line analysis, the \hi\ data cube from the VLBI data was made using {\tt ROBUST=0} weighting and 
has the same beam size as mentioned above.      
The \hi\ cube, at the full spectral resolution of $\sim$0.9\,\kms, has an rms of $\sim$0.7\,mJy\,beam$^{-1}$. 
In the spectrum of the quasar, we detect 21-cm absorption at the frequency expected from the 
GMRT and WSRT datasets.  

\section[]{Results and Discussion}
\label{sec:disc}


\subsection{Redshift of the quasar  and line of sight reddening}
\label{sec:zqso}

The IGO spectrum extracted from slit orientation 2 at the location of the quasar 
(see Fig.~\ref{overlay}) is shown in the left panel of Fig~\ref{qspec}.  
The spectrum shows two strong emission lines which we identify as Mg~{\sc ii} 
and [C~{\sc iii}] emission at \zem = 1.5266 +/- 0.0032. 
In the figure, we also overlay the redshifted SDSS composite quasar spectrum from 
\citet{VandenBerk01}.  It is interesting to note that the composite spectrum 
roughly reproduces the IGO spectrum of the quasar.    
This suggests that the line of sight reddening  towards the quasar is limited.

In order to see how the relative colors of the quasar compare
with a typical quasar at a similar redshift, we gathered colour information
of 3000 quasars in the SDSS database with 1.525$\le$\zem$\le$1.575.
Following \citet{Richards03}, we define $\Delta$(g-i) as the relative colour
of individual quasars with respect to the measured median. 
The distribution of $\Delta$(g-i) for all
these quasars is shown in the right panel of Fig.~\ref{qspec}. The measured
$\Delta$(g-i) for the quasar is $-$0.05$\pm$0.02. This
implies that this quasar is actually slightly bluer than a typical quasar at \zem = 1.52. 
Thus, the quasar sight line is relatively dust free even though it is passing 
very close to the spiral arm  of a galaxy.

In Fig.~\ref{qspec}, we also mark the expected positions of Ca~{\sc ii} and Na~{\sc i} 
absorption lines from the foreground galaxy, \ugc\,(SDSS J124357.17+404346.1). We do not 
detect these lines in the spectrum. However, this is not surprising given the poor signal 
to noise ratio and the spectral resolution (i.e $\sim 350$ \kms) that are not ideal for 
detecting these absorption lines unless they are abnormally strong. 

\subsection{SFR and metallicity of the foreground galaxy}
%
%
We use H$\alpha$, H$\beta$, [N~{\sc ii}] and [O~{\sc iii}] emission lines from the foreground galaxy, 
 detected in the IGO long-slit observations to derive the star-formation rate and metallicity.
For these, the spectra were extracted from sub-apertures of sizes $1.2\times0.5$\,kpc$^{-2}$ 
($3.5^{\prime\prime}$$\times$$1.5^{\prime\prime}$) and $0.6\times0.5$\,kpc$^{-2}$ 
($1.8^{\prime\prime}$$\times$$1.5^{\prime\prime}$)
along the slit orientations 1 and 2 respectively (see Fig.~\ref{overlay}), 
and were fitted with Gaussians to determine line fluxes and velocities.

\begin{figure*}
\centering
\includegraphics[width=150mm, angle=0]{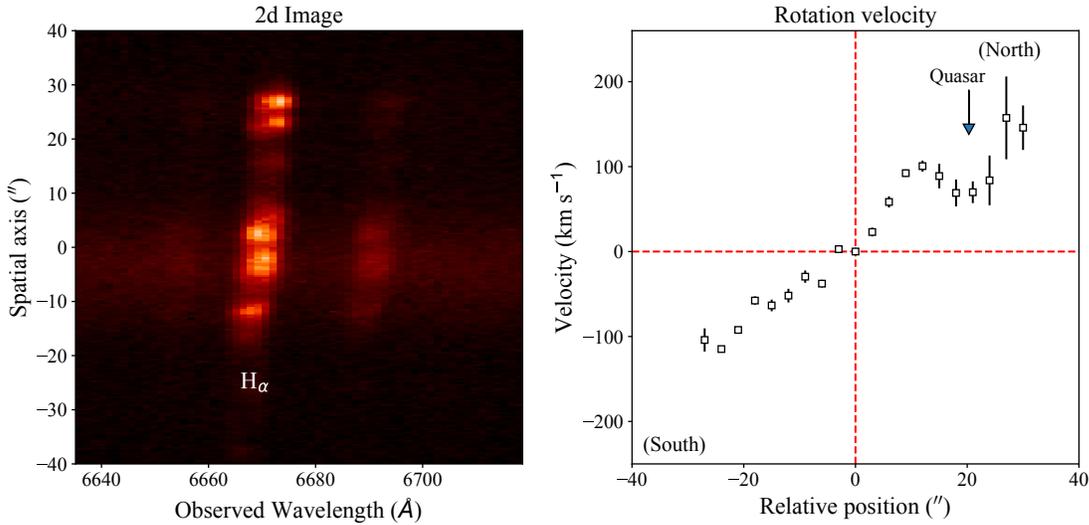}
\caption{2D spectrum and rotation velocity for H$\alpha$ from the foreground galaxy along slit orientation 1.  
}
\label{rotcur}
\end{figure*}

The 2D spectrum for the H$\alpha$ line along the slit orientation 1 and the corresponding 
rotation velocity are shown in Fig.~\ref{rotcur}. 
From the rotation velocity, it is clear that the northern spiral arm, with respect to the galactic 
center, is moving away from us. 
Following \citet{Argence09}, we estimate 
the optical depth at the intrinsic V-band of the galaxy, $\tau_V^{\rm Balmer}$, for each 
sub-aperture, using
\begin{equation}
{\rm
\tau_V^{Balmer} = \dfrac{ln\bigg(\dfrac{H\beta}{H\alpha}\bigg)- ln\bigg(\dfrac{H\beta^i}{H\alpha^i}\bigg)}{\dfrac{\tau_\beta}{\tau_V}-\dfrac{\tau_\alpha}{\tau_V}}
},
\end{equation}
an intrinsic Balmer ratio ${H\beta^i / H\alpha^i} = 2.85$ \citep{Osterbrock06} and,  
\begin{equation}
\frac{\tau_{\lambda}}{\tau_V}=\big(1-\mu\big) \bigg(\frac{\lambda}{5500\AA}\bigg)^{-1.3}+\mu\bigg(\frac{\lambda}{5500\AA}\bigg)^{-0.7},
\end{equation}
as given by \citet{Wild07a}.
Here $\tau_V$ is the total effective optical depth in V-band and $\mu$ is
the fraction of total $\tau_V$  caused by the ambient ISM. We set $\mu$ = 0.3
based on observed relations between UV continuum slope and H$\alpha$ to H$\beta$ 
emission line ratios.  The $\tau_V^{\rm Balmer}$ along the disk ranges from 
0.1 to 4.2, and the values are larger towards the galactic center compared to the disk. 
Also, within allowed errors the measured $\tau_v^{\rm Balmer}$ values are
consistent with minimal extinction in the outer spiral arms.  This is
consistent with the limited reddening seen towards the quasar J1243+4043 in Section~\ref{sec:zqso}.

Next, we estimate the star-formation rate (SFR) in each sub-aperture using, 
\begin{equation}
\log(\rm {SFR})=0.95*\log L({H\alpha}) - \log\eta{H\alpha}.
\label{equ 4}
\end{equation}
with $\log\eta{{\rm H\alpha}}= 39.38$ as given by \citet{Argence09}.  
The SFR values along  slit orientation 1 range from 0.001 to 0.4\,M$_\odot$yr$^{-1}$. 
Similarly to the dust extinction, i.e. $\tau_V^{\rm Balmer}$, we find that the maximum values of SFR 
are also seen towards the galactic center. 

Furthermore, using emission line flux of [NII]{$\lambda$}\,6583 and {\Ha} and the LINER relation given by  
\citet{Pettini04a}, we get [O/H] in individual sub-apertures.  We estimate that the metallicity is close to 
Solar (Z$_\odot$) at the center of the galaxy, and as is generally observed, decreases radially outwards. 

Since, the main objective of our analysis is to connect the gas seen in absorption 
towards the quasar with the properties of foreground galaxies, we use emission lines detected along  
slit orientation 2 to estimate the SFR and metallicity in the absorbing gas.   
Along slit orientation 2, no H$\alpha$ emission from the galaxy is detected at the 
location of the quasar.  The H$\alpha$ flux measured in the neighbouring sub-aperture suggests 
that the surface SFR is about 0.01\,M$_\odot$\,yr$^{-1}$\,kpc$^{-2}$. We take this as an upper 
limit on the SFR at the location of the quasar slight line.  
Furthermore, we find that the sub-apertures along  slit position 2 have a nearly uniform metallicity. 
The metallicity in the vicinity of the quasar line of sight is 0.5\,Z$_\odot$. 

Thus, the quasar sight line passes through a region with moderate star-formation, high metallcity, 
and very low extinction.

\subsection{Detection of \hi\ 21-cm absorption i.e. DLA associated with QGP J1243+4043}

\begin{figure}
\hbox{
\includegraphics[width=80mm]{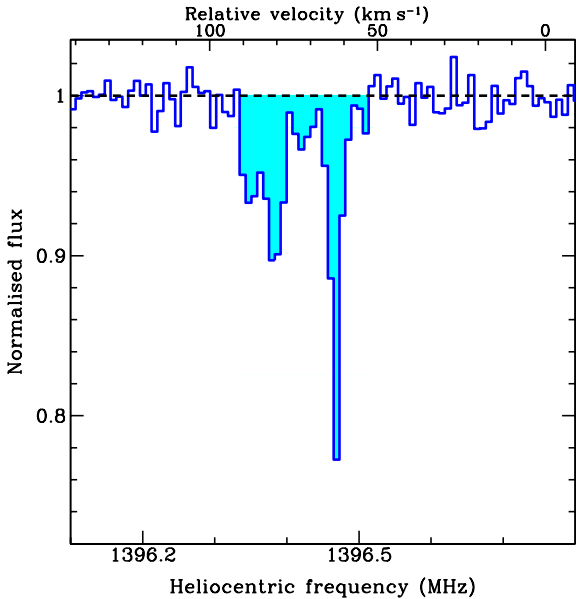}
}
\caption{
GMRT 21-cm absorption spectrum towards the quasar J1243+4043.  
The spectral resolution and rms are $\sim$1.7\,\kms\ and 1.8\,mJy\,beam$^{-1}$\,channel$^{-1}$, 
respectively. 
The zero of the velocity scale is defined at $z$=0.01693, the redshift of \ugc.  
}
\label{gmrt}
\end{figure}
\begin{figure}
\vbox{
\includegraphics[trim = {1cm 0 0 0}, clip, width=75mm, angle=00]{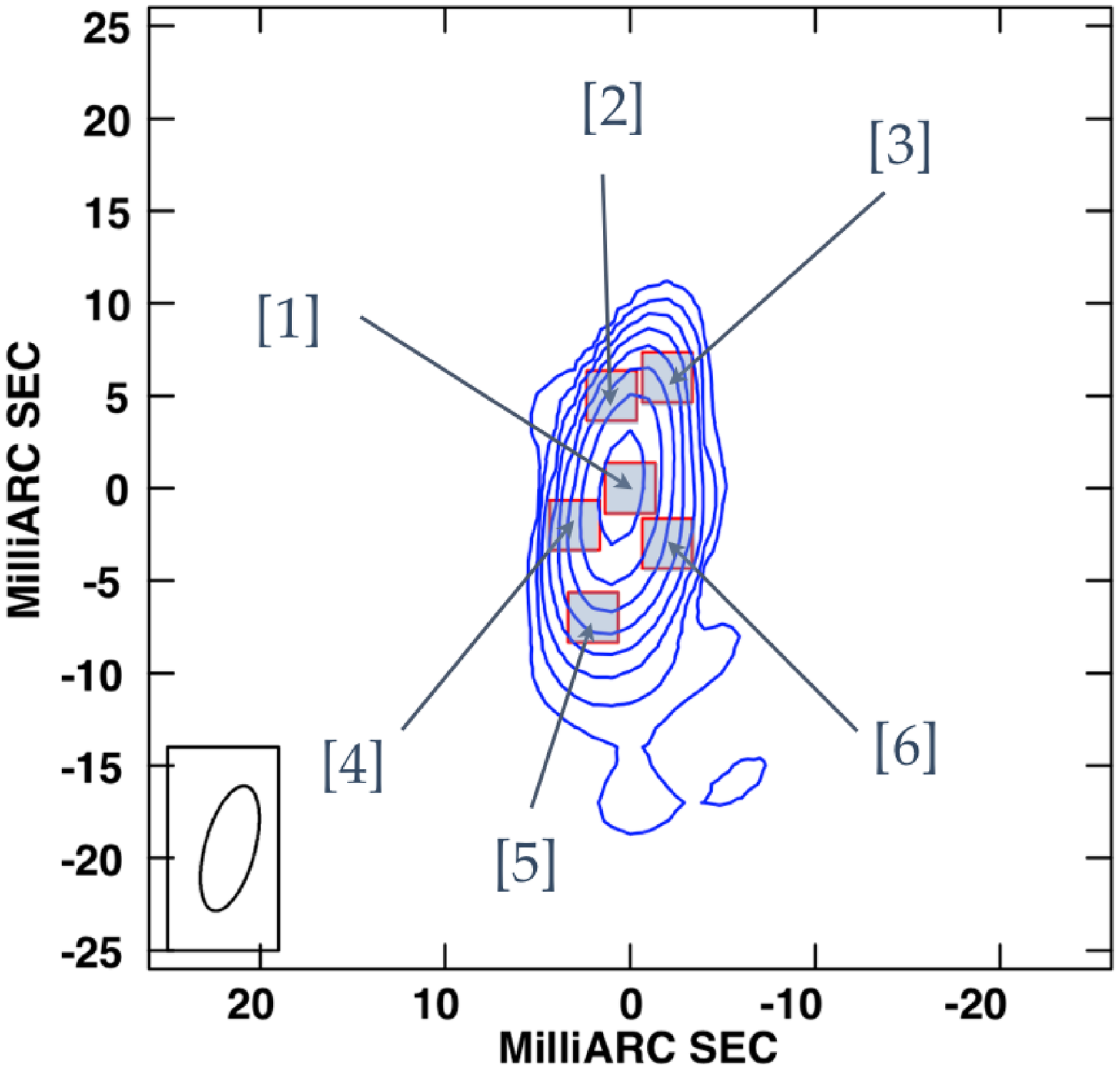}
\includegraphics[width=75mm]{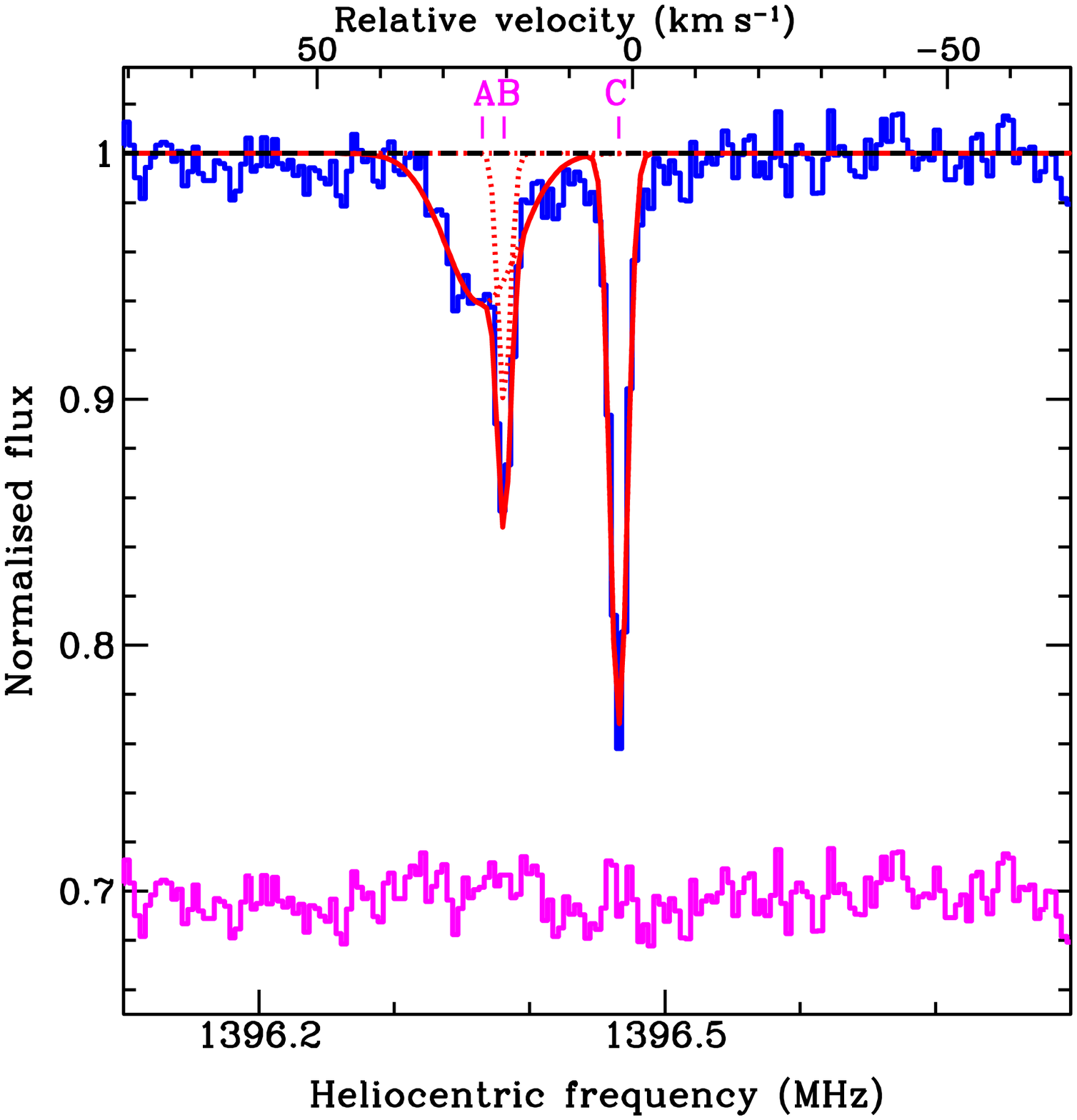}
}
\caption{
VLBI image ({\tt ROBUST = 0}) centered at the core ($\alpha$(J2000)=12:43:55.78142, $\delta$(J2000)=40:43:58.4440; {\it upper panel})  
and the full resolution ($\sim$0.9\,\kms) VLBI spectrum towards the core ({\it lower panel}). 
Contour levels in the radio continuum image are 0.5$\times$(-1, 1, 2, 4, 8, ...)\,mJy\,beam$^{-1}$.  
The restoring beam of 7.0$\times$2.7\,mas$^2$ (position angle $\sim$-15$^\circ$) 
is shown as an ellipse. 
Individual Gaussian components centered at A, B and C (see Table~\ref{21cmtable}), and the resulting fits to the 
spectrum are plotted as dotted and continuous lines, 
respectively.  The residuals, on an offset arbitrarily shifted for clarity, are also shown.  
The zero on the velocity scale corresponds to the 21-cm absorption peak. 
The labels [1] to [6] mark regions used to investigate the variation of 21-cm optical depth 
across the radio source (see also Fig.~\ref{vlbispec}). 
}
\label{vlbi}
\end{figure}

\citet{Dutta17} previously reported the \hi\ 21-cm absorption towards \ugc. 
In Fig.~\ref{gmrt}, we present our higher resolution GMRT spectrum. 
With respect to the emission line redshift ($z_g$=0.01693$\pm$0.00001) based on the SDSS spectrum 
corresponding to the center of the galaxy, the absorption peak is redshifted by 
$\sim$65\,\kms (see Fig.~\ref{gmrt}).
The absorption is also detected in the VLBI and WSRT data. These spectra are presented in 
subsequent sections.     

Even though H$\alpha$ emission is not detected along the line of sight to the quasar,
the 21-cm absorption has a velocity range very similar to the H$\alpha$ emitting gas 
in the disk at this galactocentric radius (see Fig.~\ref{rotcur}). Thus, it appears that 
the 21-cm absorption in this case originates from the gas co-rotating with the stellar disk.

In the GMRT spectrum (see Fig.~\ref{gmrt}), 90\% of the total 21-cm optical depth is contained 
within 30\,\kms\, and the total integrated 21-cm optical depth, 
$\int \tau (v) dv = 2.24\pm0.10$. 
For an optically thin cloud the integrated 21-cm optical depth is related to the neutral 
hydrogen column density $N$(H~{\sc i}), spin temperature $T_{\rm s}$, and covering factor $f_c$ through,
\begin{equation}
N{(\ion{H}{i})}=1.823\times10^{18}~{T_{\rm s}\over f_{\rm c}}\int~\tau(v)~{\rm d}v~{\rm cm^{-2}}.
\label{eq1}
\end{equation}
For $f_c =$1, as is the case for this absorber (see Section~\ref{sec:vlbi}) and adopting $T_{\rm s}=$70\,K, 
which is the median column density weighted $T_{\rm s}$ for the cold neutral medium (CNM) in our 
Galaxy \citep[][]{Heiles03}, we get $N$(\hi) = 2.9$\times10^{20}$($T_{\rm s}$/70)(1.0/$f_c$)\,cm$^{-2}$. 
Thus, for temperatures typically seen in the CNM gas in the Galaxy, the 21-cm absorber detected 
towards quasar J1243+4043 can be classified as a DLA.

\subsection{Parsec scale structure in cold atomic gas}
\label{sec:vlbi}

\begin{figure*}
\vbox{
\hbox{
\includegraphics[width=60mm, angle=0]{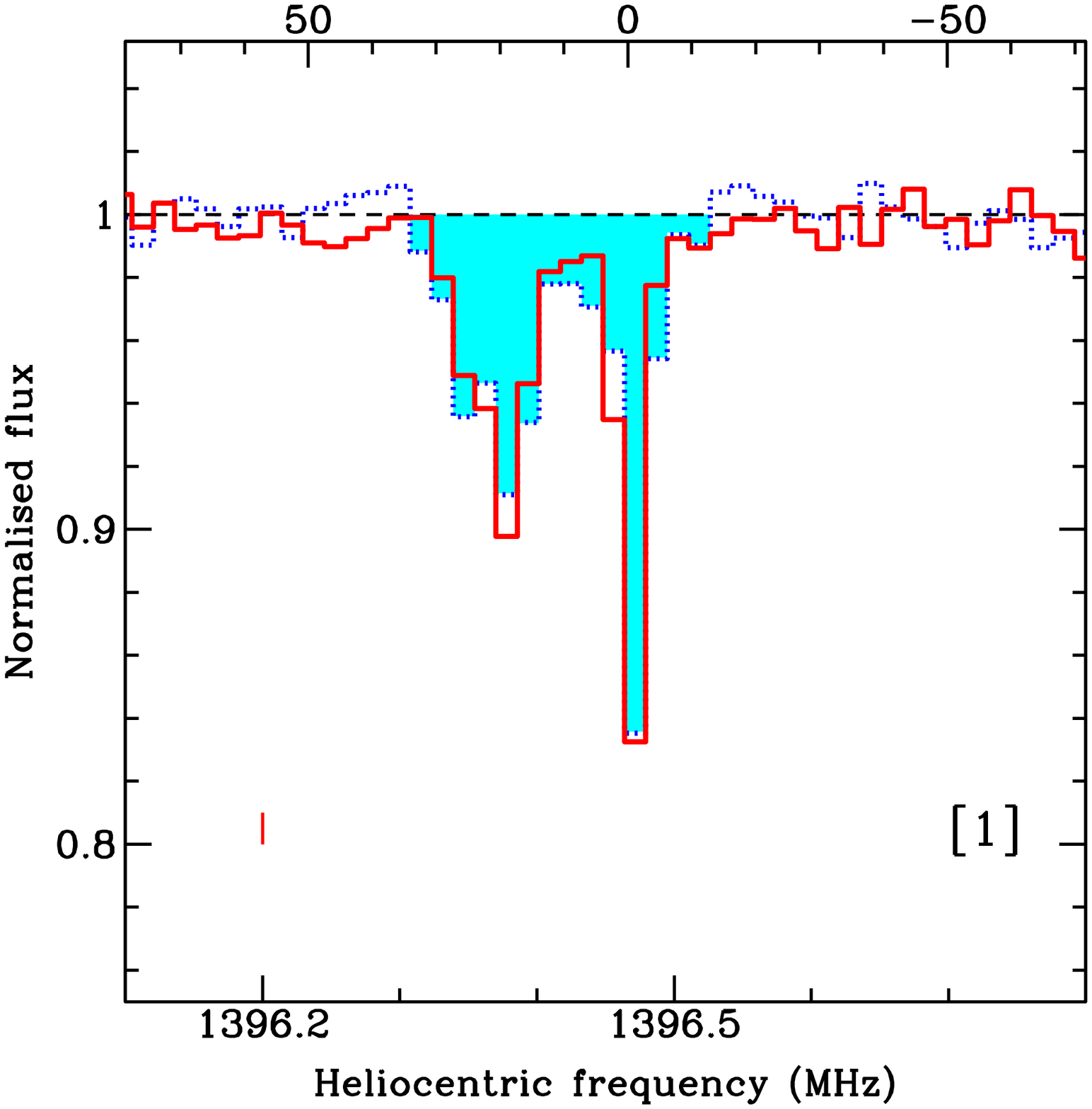}
\includegraphics[width=60mm, angle=0]{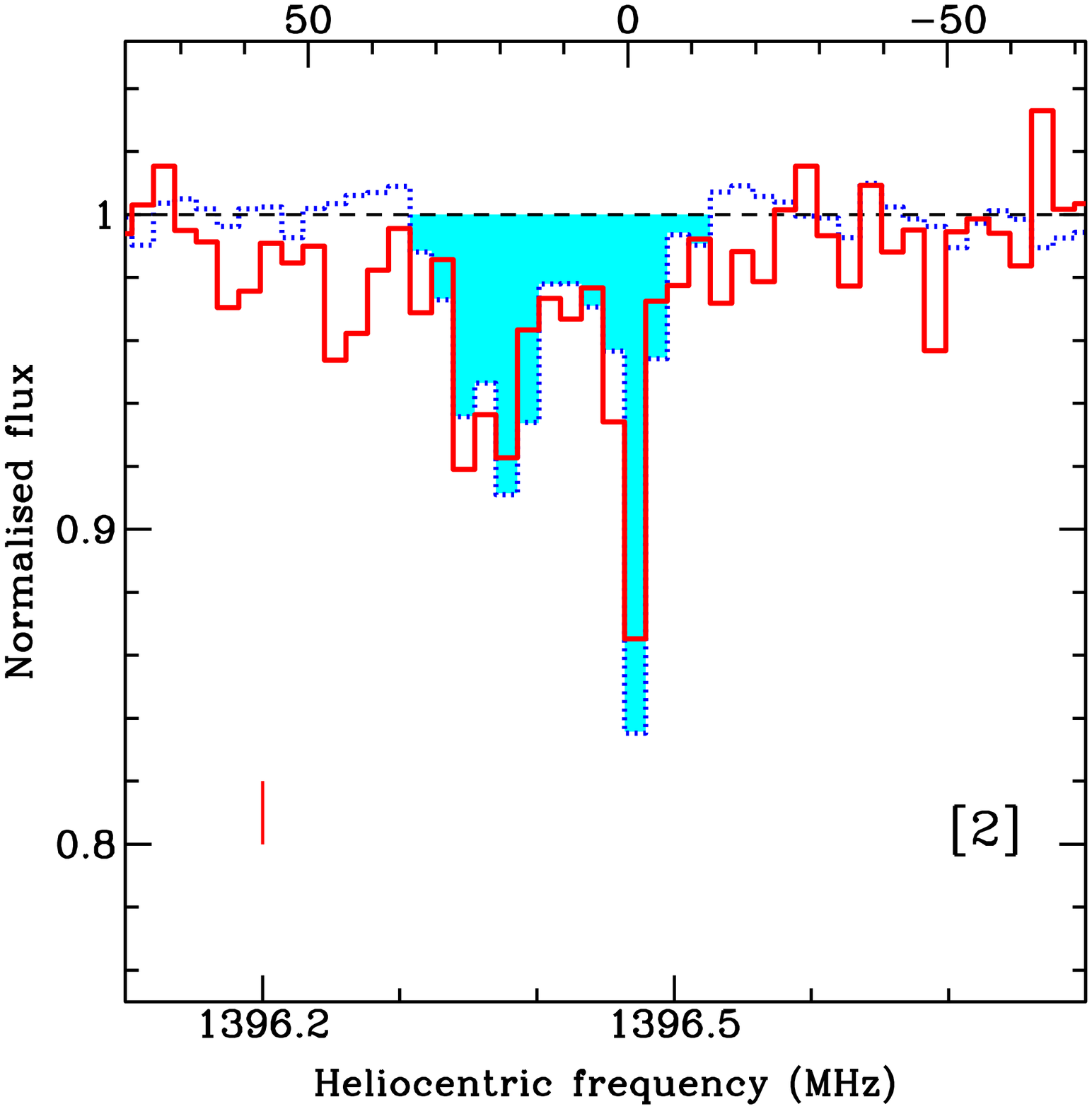}
\includegraphics[width=60mm, angle=0]{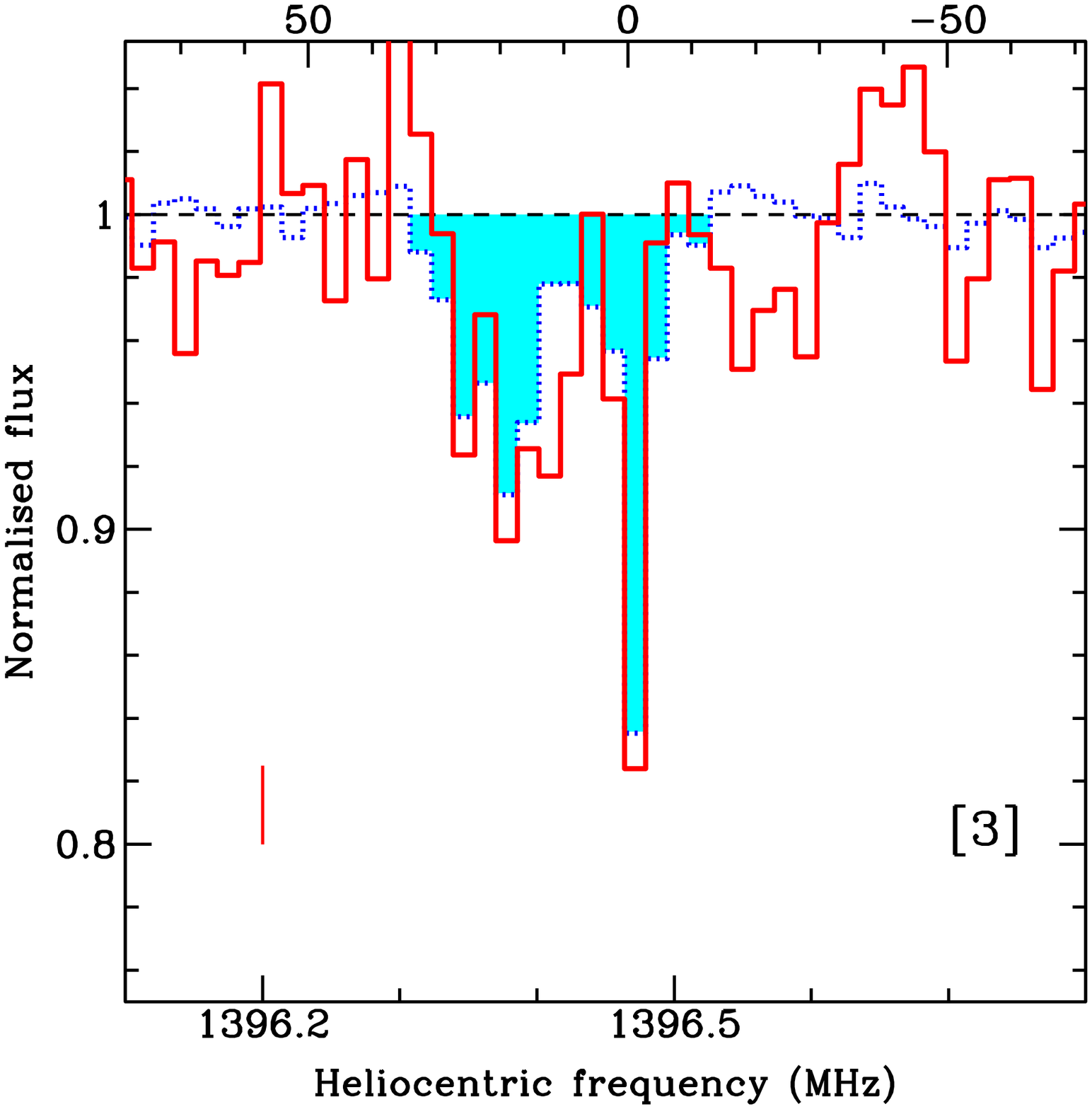}
}
\hbox{
\includegraphics[width=60mm, angle=0]{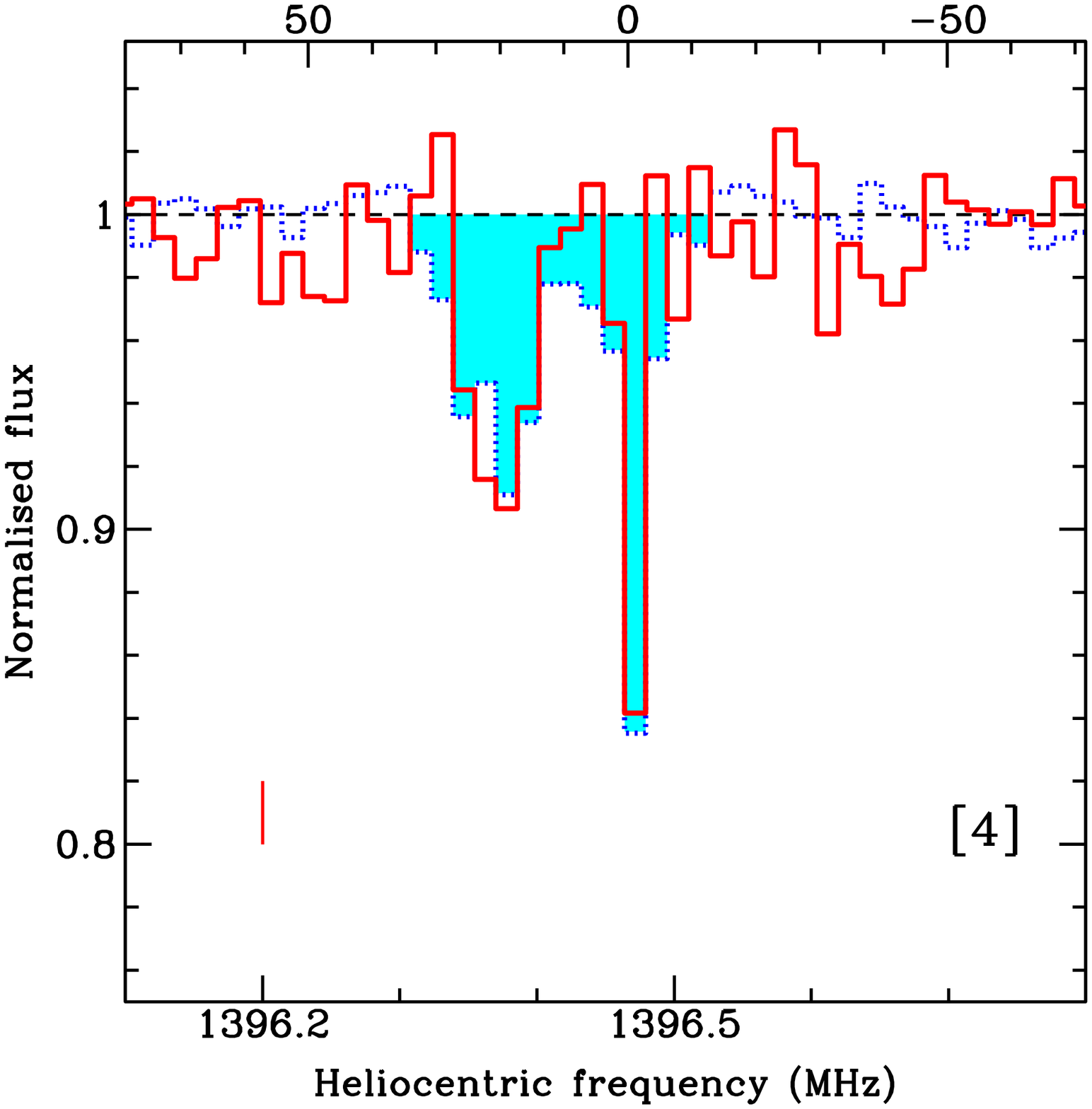}
\includegraphics[width=60mm, angle=0]{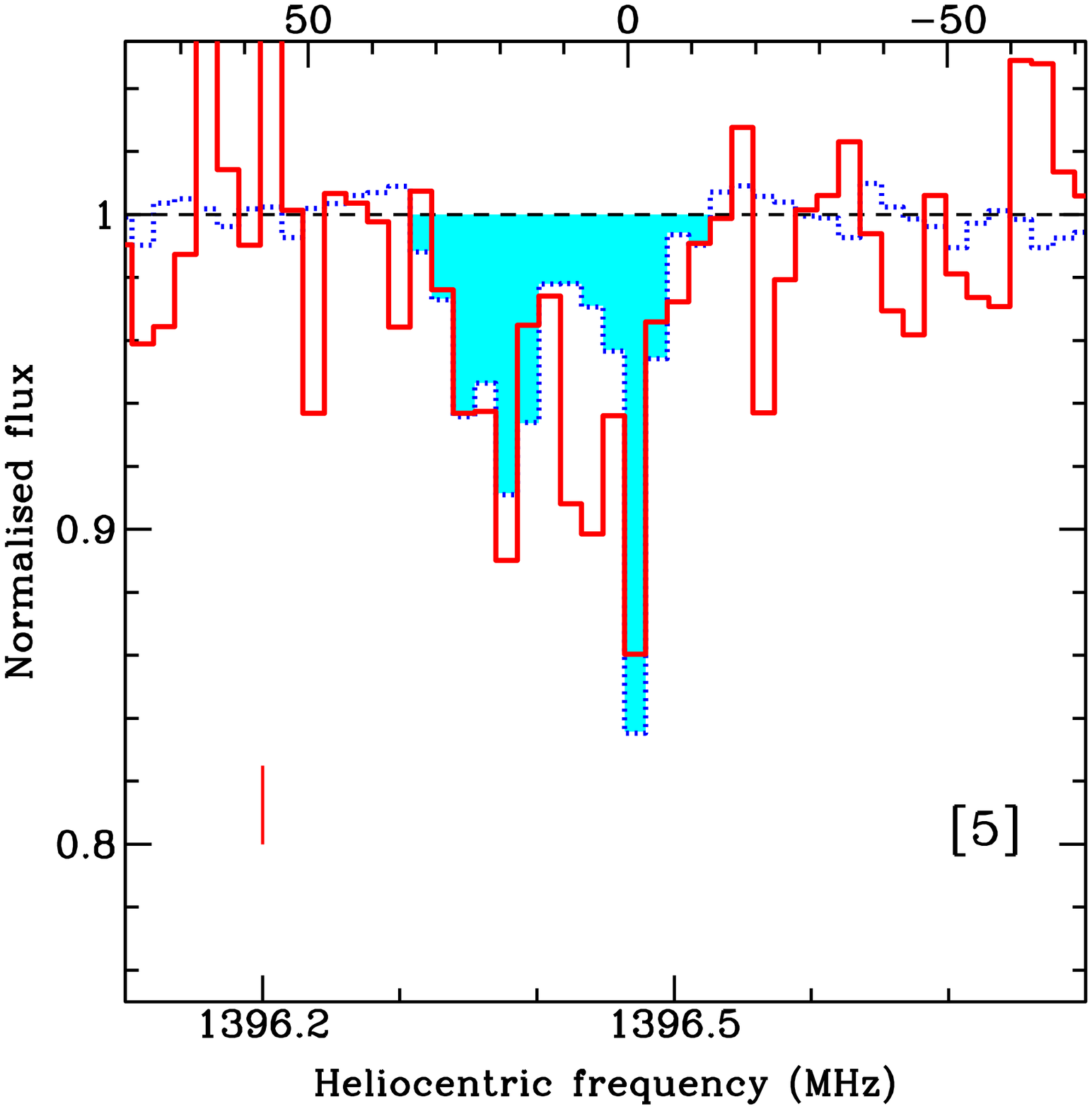}
\includegraphics[width=60mm, angle=0]{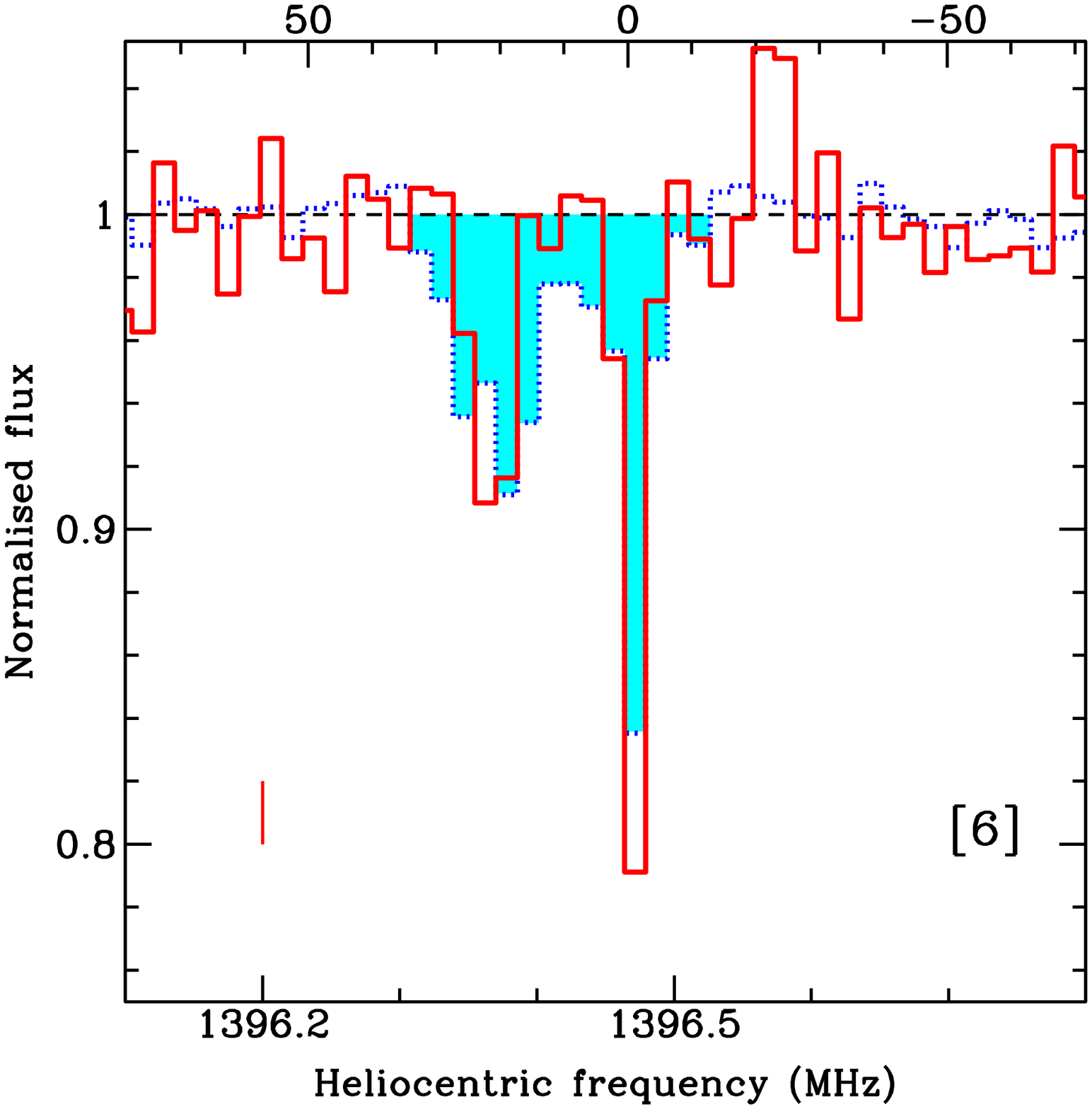}
}
}
\caption{
VLBI 21-cm absorption spectra (solid line) extracted from regions marked as [1] to [6] in Fig.~\ref{vlbi}. 
For reference, the GMRT spectrum (dotted line/ shaded) assuming full coverage of the arcsecond-scale radio emission is also plotted.
Both the VLBI and GMRT spectra in these panels have been smoothed to 4\,\kms\ and resampled on to the 
same frequency scale. The velocity scale (top axis) is same as in Fig.~\ref{vlbi} ({\it lower panel}).  
The vertical line in each panel represents the 1$\sigma$ error in the corresponding VLBI spectrum.   
}
\label{vlbispec}
\end{figure*}
 
The VLBI image ({\tt ROBUST = 0}) of the quasar at $\sim$1396\,MHz, is shown in the {\it top} 
panel of Fig.~\ref{vlbi}.     
The quasar exhibits a core-jet morphology with the jet extending to the South with an 
overall extent of 25\,mas, i.e. $\sim$9\,pc at the redshift of the foreground galaxy.  
The morphology of the northern and southern components is better revealed 
in the higher resolution (3.0\,mas$\times$1.8\,mas) 
5\,GHz image from the VLBA Imaging and Polarization Survey \citep[VIPS; ][]{Helmboldt07}.  
About 99\% of the total cleaned flux density (165\,mJy) in 5\,GHz image can be 
modeled with two Gaussian components separated by 3.6\,mas (i.e. 1.2\,pc at the $z_g$), 
and having integrated flux densities of 105\,mJy and 58\,mJy in the northern and southern components respectively. 
In our $\sim$1396\,MHz image, the northern component, identified as core on the 
basis of flatter spectral index and hence coincident with the optical quasar,  has a peak 
flux density of 88\,mJy\,beam$^{-1}$.  
The contour plot shown in the middle panel of Fig.~\ref{vlbi} is centered at this core 
component.  
The total {\tt CLEAN}ed flux density of the quasar in this image is $\sim$140\,mJy. 
The total {\tt CLEAN}ed flux density in a lower spatial resolution image (not shown here) made with 
{\tt Natural} weighting is 167\,mJy.  
The diffuse radio emission in this image extends up to 100\,mas (35\,pc at the $z_g$) southward from the `core'.  
From the WSRT interferometric data obtained at the same epoch, 
we measure the total flux density of the quasar at arcsecond scales to be $\sim$198\,mJy.
Thus, at VLBI scales we recover $\sim$85\% of the total arcsec scale flux density. 

In the {\it bottom} panel of Fig.~\ref{vlbi}, we also show the full resolution ($\sim$0.9\,km\,s$^{-1}$) 
21-cm absorption spectrum (rms$\sim$0.7\,mJy\,beam$^{-1}$\,channel$^{-1}$)  towards 
the radio continuum peak.  
To investigate the variation of the 21-cm optical depth across the radio source, 
we define six regions over the extent of the radio source ($\sim$15\,mas).  
These regions, labelled as [1] to [6] in Fig.~\ref{vlbi}, have sizes of 3\,mas$\times$3\,mas. 
The normalized 21-cm absorption spectra corresponding to these are shown 
in Fig.~\ref{vlbispec}.  As the continuum is lower in regions beyond the peak,  
to enhance the signal-to-noise ratio the spectra have been smoothed to 4\,\kms.  
The spectrum from region [1] has the maximum optical depth sensitivity ($\tau_{3\sigma}=$ 0.02).  
The sensitivity ($\tau_{3\sigma}=$ 0.08) is minimum for region [5] and is inadequate 
beyond (southward) it to detect the absorption.

Two conclusions can be drawn from the comparison of VLBI spectra from different regions: 
{\it (i)} the 21-cm absorption profiles are similar across (15\,mas$\sim$5\,pc) the radio source. 
The largest detectable difference is between regions [1] and [5] at 5\,\kms\, and it is significant 
only at 2.5$\sigma$, and
{\it (ii)} the parsec-scale absorption spectra are also consistent with the GMRT 21-cm absorption spectrum 
(also shown in Fig.~\ref{vlbispec}).  
The former implies that the size of the absorbing clouds A, B and C are $\ge$5\,pc, i.e. the extent 
of the radio source. The latter under the plane-parallel slab approximation imply that the clouds also cover 
the remaining $\sim$30\% of the radio continuum emission resolved out in the {\tt ROBUST =0} VLBI image.  
If the clouds covered only the radio emission detected in Fig.~\ref{vlbi} then under the plane-parallel slab 
approximation the normalized flux densities for GMRT and VLBI spectra would have been different by 6$\sigma$.  
Therefore, based on the extent of radio emission detected in the VLBI image with {\tt Natural} weighting, 
we conclude that the size of absorbing clouds $\ge35$\,pc.

The extent of the absorbing gas observed here could be a direct consequence of coherent 
structures present in the diffuse atomic gas in galaxies. 
Similar cloud sizes have been inferred from the VLBI spectroscopic studies that have been  possible 
for a handful of low-$z$ \hi\ 21-cm absorbers.  
In the case of the $z_{abs}$ = 0.0912 DLA towards B0738+313, \citet[][]{Lane00} found the background
source to be partially resolved at mas scales. Within the measurement uncertainties they do not 
find any strong variations in the \hi\ optical depth across 20\,pc.  
In the case of QGP 3C\,232 $-$ NGC\,3067 no 21-cm optical depth variations have been seen across 
$\sim$2-20\,pc \cite[][]{Keeney05}.  
For the $z$ = 0.03321 galaxy towards J104257.58+074850.5, \citet[][]{Borthakur11} found that the 21-cm 
absorption is  similar over 27.1\,pc$\times$13.9\,pc \cite[see also][]{Dutta16, Allison16}.
Similarly, \citet{Gupta12} used VLBA continuum images of 52 quasars with \hi\ 21-cm absorption optical depth 
measurements at 0.5$<z<$1.5 to conclude that the 
21-cm absorbing gas is patchy and has a typical correlation length of 30-100\,pc 
\citep[see also][]{Braun12, Curran13}. 

At $z\gapp$2 where large samples of DLAs are available, \hi\ 21-cm and Ly$\alpha$ absorption 
spectra towards radio bright quasars can be combined to determine the spin temperature, T$_{\rm s}$, of the gas.  As VLBI spectroscopy is 
not possible at high redshifts, these studies use the core fraction (i.e. the ratio of flux density 
detected in the mas- and arcsecond-scale images) to correct the optical depth for partial coverage 
using a single value of covering factor, $f_c$, and to measure T$_{\rm s}$ 
\citep[e.g.][]{Kanekar09vlba, Srianand12, Kanekar14}.
The typical upper limit on the extent of radio emission from these observations is $\sim$300\,pc.  
This is consistent with the lower limits on the sizes of absorbing clouds 
inferred from low-$z$ VLBI spectroscopic observations and justifies the practice of using a single 
covering factor to correct for the partial coverage of radio emission.  
However, we caution that this assumption may only be valid for current samples of absorbers 
dominated by quasar sight lines tracing diffuse atomic gas with little dust extinction 
\citep[color excess, E(B-V)$<$0.001 to 0.085][]{York06}.  
\hi\ 21-cm optical depth variations of the order of a few over 10-100\,pc have been observed 
for sight lines towards reddened quasars tracing denser ISM phases \citep[][]{Srianand13dib, Biggs16}.

\subsection{Spin temperature of the absorbing gas}
\label{sec:temp}
%
%
\begin{figure}
\hbox{
\includegraphics[trim=0cm 10.0cm 0.0cm 0.0cm, clip=true, width=85mm]{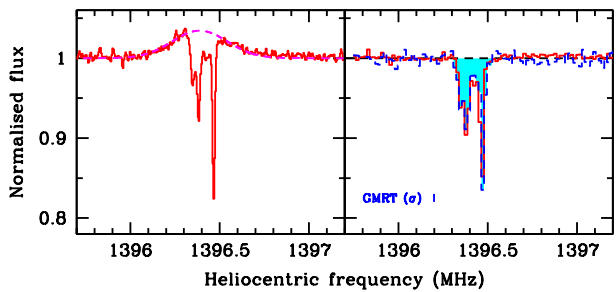}
}
\caption{
{\it Left:} 
The WSRT spectrum towards quasar J1243+4043.  A single Gaussian fit to the \hi\ emission 
is plotted as a dashed line.   
{\it Right:} 
The WSRT spectrum obtained after subtracting the Gaussian model fitted to the \hi\ emission, 
and the GMRT spectrum (shaded profile) smoothed to 4\,\kms.  The 1$\sigma$ error for the GMRT spectrum is also shown.   
}
\label{emabs}
\end{figure}

Our radio and optical data allow us to constrain the temperature of the absorbing gas 
using several methods.
From the WSRT {\tt ROBUST=0} \hi\ cube, the spectrum extracted towards the quasar  
exhibits both 21-cm emission and absorption ({\it left} panel of Fig.~\ref{emabs}; 
see also Section~\ref{sec:kin}).  
We model the \hi\ emission towards the quasar by fitting a single Gaussian of FWHM 79\,\kms.  
The pixels with \hi\ absorption were masked during this process. 
As a consistency check, we subtracted this Gaussian model from the WSRT profile.  
The resulting absorption spectrum, which is consistent with the absorption detected in 
the GMRT spectrum, is shown in the {\it right} panel of Fig.~\ref{emabs}.    

Assuming optically thin emission, the \hi\ line intensity can be converted 
to an \hi\ column density using,
\begin{equation}
N{(\ion{H}{i})}=\frac{1.104\times10^{21}}{b_{\rm maj}\times b_{\rm min}}~\int~S(v)~{\rm d}v~{\rm cm^{-2}}, 
\label{eqhiem}
\end{equation}
where $S(v)$ is in mJy\,beam$^{-1}$\,\kms, $v$ in \kms, and $b_{\rm maj}$ and $b_{\rm min}$ are 
the beam major and minor axes in arcsec.  
For a single Gaussian model fitted to the \hi\ emission profile, this yields  
$N$(\hi)=1.6$\times$10$^{21}$\,cm$^{-2}$. Using this and the integrated 21-cm 
optical depth from the GMRT spectrum with $f_c$=1 in Equation~\ref{eq1}, we get 
a harmonic mean spin temperature of 390\,K.    

Alternatively, we can also use the measured surface SFR and the  
Kennicutt-Schmidt law to estimate the \hi\ column density along the quasar sight line.
No H$\alpha$ emission from the galaxy is detected at the location of the quasar. 
The SFR measured in the immediate neighborhood suggests that the surface 
SFR$\sim$0.01\,M$_\odot$\,yr$^{-1}$ kpc$^{-2}$. We take this as an upper limit on 
the SFR at the location of the quasar sight line.
The Kennicutt-Schmidt law is given by,  
\begin{eqnarray}
<\dot{\psi_\star}>_\perp & = & 0, ~{\rm when} ~N_\perp < N_\perp^{\rm crit}\nonumber,\\
                         & = &  K (N_\perp/N_\perp^{\rm c})^\beta,~~~ N_\perp\ge N_\perp^{\rm crit}.
\label{sklaw}
\end{eqnarray}
where $K=(2.5\pm0.5)\times10^{-4}$~M$_\odot$~yr$^{-1}$, $\beta = 1.4\pm0.15$, 
$N_\perp^{\rm c} = 1.25\times 10^{20}$~cm$^{-2}$ and log~$N_\perp^{\rm crit}$ = 20.62  
\citep{Kennicutt98, Kennicutt98b}. 
Using the above mentioned SFR, we get $N$(H~{\sc i}) $\le 1.7 \times 10^{21}$ cm$^{-2}$ 
along the quasar sight line.  
This is consistent with the \hi\ column density measured from the \hi\ emission profile, 
and corresponds to a harmonic mean spin temperature of $T_{\rm s}\le$410\,K. 

\begin{table}
\caption{Multiple Gaussian fits to the 21-cm absorption profile towards the radio core at 
mas scales.}
\begin{tabular}{cccc}
\hline
{\large \strut} Component & $z_{abs}$ & FWHM         &  $\tau_p$   \\
          &           & (\kms)       &               \\
\hline
{\large \strut} A & 0.017217 & 13.7$\pm0.9$ &0.061$\pm0.004$     \\
B & 0.017205 & 2.3$\pm0.2$  &0.108$\pm0.009$     \\
C & 0.017143 & 3.1$\pm0.1$  &0.264$\pm0.007$     \\
\hline
\end{tabular}

\label{21cmtable}
\end{table}

\begin{figure*}
\hbox{
\includegraphics[width=85mm, angle=0]{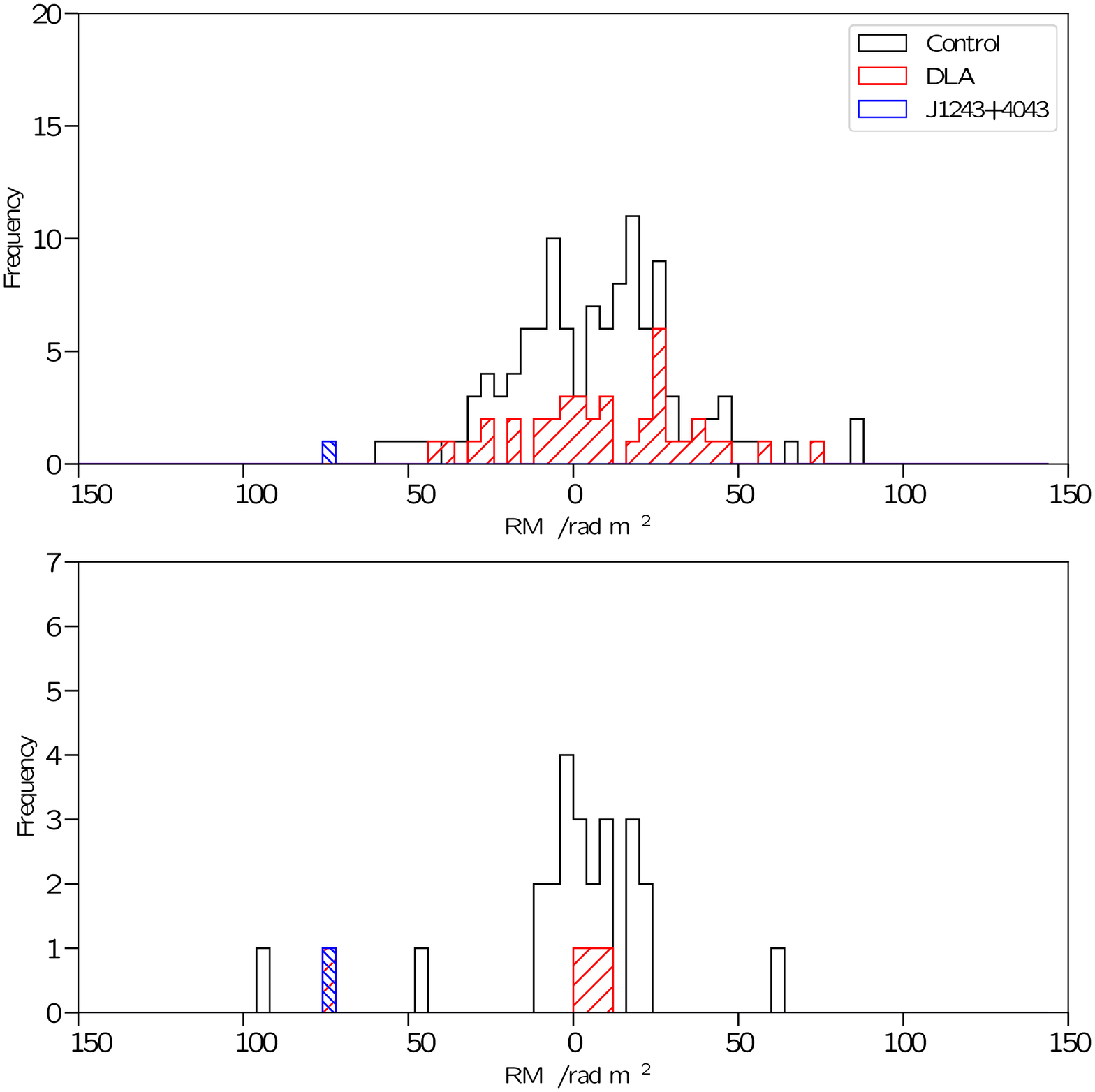}
\includegraphics[width=85mm, angle=0]{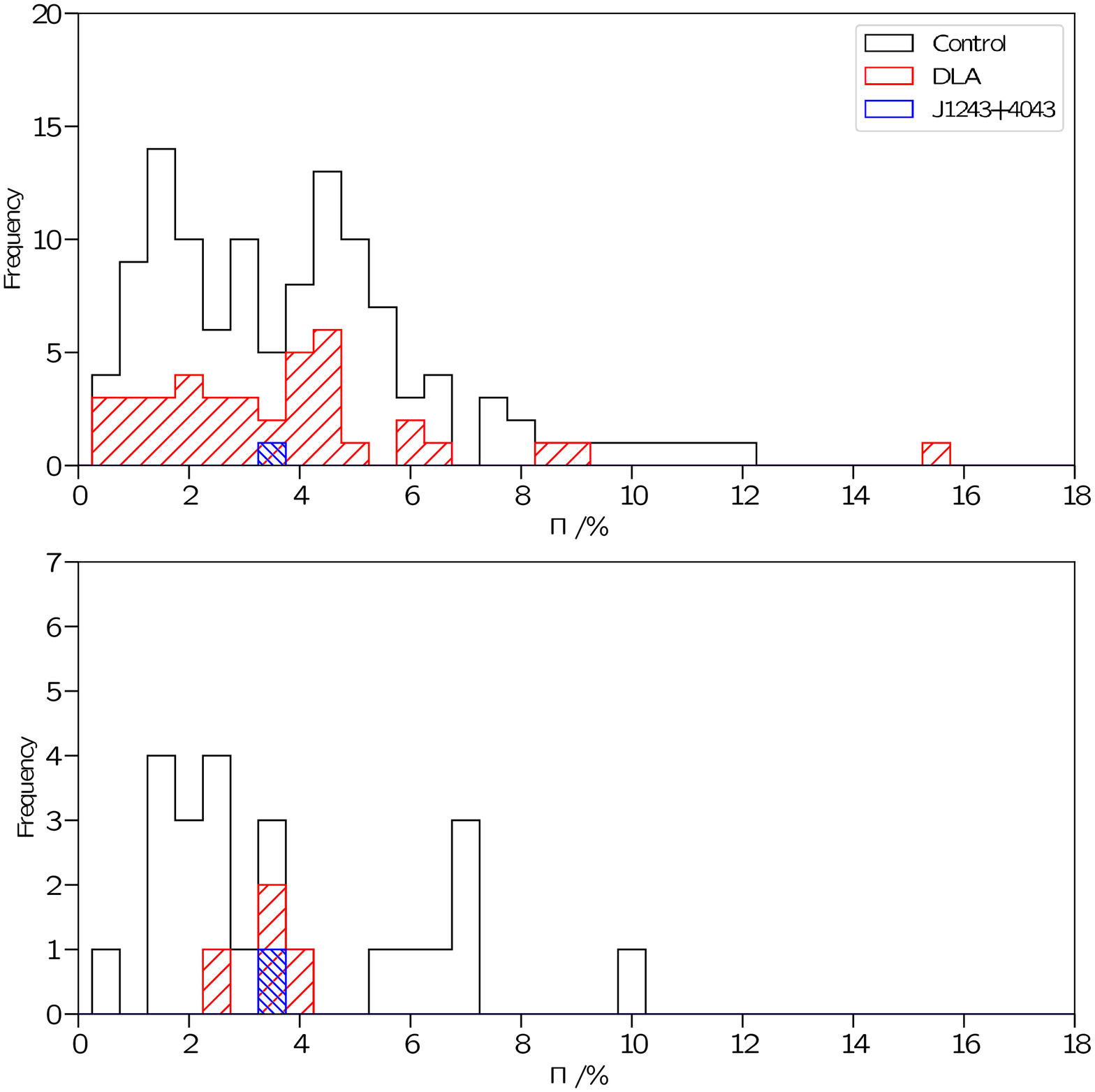}
}
\caption{
The distribution of RM ({\it left})  and polarization fraction ({\it right}) for high-$z$ DLAs (top panels) 
and low-$z$ 21-cm absorbers (bottom panels).    
The control refers to sight lines with no DLA ({\it top}) or no 21-cm absorption ({\it bottom}), respectively.   
}
\label{RMPIdist}
\end{figure*}

%
For a simple two-phase medium the harmonic mean spin temperature of atomic gas is related 
to the spin temperatures of the CNM and Warm Neutral Medium (WNM) phases through the following equation,
\begin{equation}
\frac{1}{T_{\rm s}} = \frac{f_{\rm CNM}}{T_{\rm s,CNM}} + \frac{1 - f_{\rm CNM}}{T_{\rm s,WNM}},
\label{eqharm}
\end{equation}
where $f_{\rm CNM}$ is the fraction of atomic gas in the CNM phase.
For the CNM and WNM in the Milky Way, the kinetic temperatures of the CNM and WNM phases are 
$\sim$40-200\,K and 5500-8500\,K, respectively \citep[][]{Wolfire95}.
It is well known that T$_{\rm s}$ depends on the local kinetic temperature of the gas, and 
in particular, T$_{\rm s}$ = T$_{\rm K}$ for CNM and T$_{\rm s} \le$ T${\rm _K}$ for 
the WNM \citep[][]{Liszt01}.  
To constrain kinetic temperature of the CNM phase towards J1243+4043, 
we model the absorption profile shown in Fig.~\ref{vlbi} using multiple Gaussian 
components.  The fitted Gaussian parameters for components A, B and C are presented 
in Table~\ref{21cmtable}.  From the width of the narrowest component B\footnote{For A and C, the 
FWHMs correspond to kinetic temperatures of 4100\,K and 210\,K, respectively.} and 
assuming that the line is purely thermally broadened, we determine 
$T_{\rm K}$ = 115\,K.  This is very close to the typical values ($\sim$100\,K) of 
kinetic temperature observed in the Milky Way \citep[e.g.][]{Heiles03}. 
We adopt 115\,K as the column density weighted harmonic mean spin temperature 
of the CNM phase, as detected in absorption.  The temperature of the WNM phase is warm enough 
that it does not produce any detectable 21-cm absorption in the spectra.   
With this simple but powerful assumption for the sight line towards 
J1243+4043 ($T_{\rm s}$ = 400\,K), we estimate $f_{\rm CNM}$ = 0.27. This is remarkably 
consistent with the median CNM fraction observed in the Galaxy \citep[e.g. Fig.\,7 of][]{Heiles03}. 
Note that the majority of DLAs at high-$z$ exhibit CNM fractions significantly less than this 
\citep[][]{Srianand12dla, Kanekar14}. 
  
We would like to caution that while we detect absorption over 5-10\,pc, the  
H$\alpha$ based SFR and the 21-cm emission are both measured over $\sim$0.5 and 7\,kpc respectively. 
The spatial inhomogeneities and structures in the gas may introduce large systematic errors 
into various total $N$(\hi) estimates.  
Therefore, estimating total $N$(\hi) and T$_{\rm s}$ towards this sight line using 
Ly$\alpha$ absorption would be of much interest.

\subsection{Faraday rotation and HI 21-cm absorption}
\label{sec:rotm}
%
%
If the background quasar is polarized then the magnetoionic plasma along the line of sight can also 
be detected via the effect of Faraday rotation, in which the polarization angle of the linearly 
polarized synchrotron emission rotates as it propagates through the intervening medium. The rotation measure (RM, in rad~m$^{-2}$) is given by:
\begin{equation}
{\rm RM} = 0.81\, \int^{s}_{0} n_e B_\parallel dl, 
\end{equation}
where $B_{\parallel}$ in $\muup$G is the component of the magnetic field that lies parallel to the 
line of sight, $n_{e}$ in cm$^{-3}$ is generally the electron density, and $dl$ in pc is an 
infinitesimal element of the path length.  
In particular, in this context the samples of polarized quasars with intervening absorbers such as Mg~{\sc ii} 
and DLAs that are known to be associated with galaxies are of great interest.  There have been several studies 
in the past to use samples of intervening absorbers to probe the cosmic evolution of magnetic fields
\citep[e.g.][]{Kronberg82, Wolfe92, Bernet08, Farnes14, Malik17}.  

The quasar J1243+4043 has an 
RM$=-71.8\pm10.0$~rad~m$^{-2}$ and a polarized fraction of $P/I=3.71\pm0.22$\% at 1.4~GHz 
\citep[][]{Taylor09}. 
The foreground is likely not a substantial issue at a Galactic latitude of $76.3^{\circ}$, and 
the catalogue of \citet[][]{Oppermann15} indicates a small Galactic Rotation Measure contribution of -5.1$\pm$4.0\,rad\,m$^{-2}$.
One could further argue that one requires a flat-spectrum radio source in order to accurately probe a 
foreground intervening medium \citep[][]{Farnes14}. As J1243+4043 shows very little flux variability 
at L-band, we combine the peak flux densities 
from various radio surveys: the TGSS-ADR at 150\,MHz of 336.98\,mJy, WENSS at 
325\,MHz of 303.98\,mJy, the NVSS at 1.4\,GHz of 171.5\,mJy, and GB6 at 4.86\,GHz of 125.81\,mJy. 
This allows us to derive an integrated spectral index of $\alpha\approx-0.3$, and so J1243+4043 
is well-suited for such a foreground study.

While the polarized fraction of J1243+4043 is rather typical of the underlying 1.4\,GHz source population, the $|$RM$|$ is higher than any other source in the high-$z$ DLA sample of \citet[][]{Farnes17}. 
Following \citet[][]{Farnes17} and assuming a normal distribution of RMs, we calculate a Bayesian probability of 88.7$_{-6.8}^{+4.5}$\% that the RM of J1243+4043 is greater than the RM of the 
DLA sample from \citet[][]{Farnes17}. The stated errors in the calculated probability 
represent the $1\sigma$ uncertainty for the RM of J1243+4043. The location of J1243+4043 relative 
to the distribution of RMs and polarized fractions in \citet[][]{Farnes17} is shown in 
Fig.~\ref{RMPIdist} (top panels). A key component of the \citet[][]{Farnes17} data are that the 
DLAs and background quasars are all located at high redshifts, with a median DLA redshift of 
2.11 and a quasar redshift of 2.48.

We therefore compare these data with a catalogue of low redshift QGPs with \hi\ 21-cm absorption 
measurements taken from the literature \citep[][Gupta et al. in prep.]{Dutta17}, and RM from 
\citet[][]{Taylor09}.  In this sample, there are 20 sightlines (excluding J1243+4043). Three of 
these are detected in \hi\ 21-cm absorption.  
This sample has a median quasar redshift of 1.2 and a foreground galaxy redshift of 0.03.  The impact 
parameters of quasar sight line from the center of galaxy are in the range: 3 - 27\,kpc 
(median: 15\,kpc).  
This QGP sample is unique compared to the DLA sample in two ways: (i) the redshift range is comparable, and 
(ii) in all the cases we know whether a sight line passes through the optical extent of galaxy or not.  
Specifically, in this sample only the 21-cm detections are associated with sight lines through optical/\hi\ 
disks of galaxies and can be classified as a DLA (for T$_{\rm s}$=100\,K).  
The location of J1243+4043 relative to the distribution of RMs and polarized fractions in this 
new catalogue where control refers to the sight lines with 21-cm non-detections is shown in the 
bottom panels of Fig.~\ref{RMPIdist}. 

Critically, there is no detectable difference between the control versus DLA or 21-cm absorber samples in either 
catalogues. The only difference between the datasets are the distributions of the high- versus low-redshift RMs. 
We calculate a Bayesian probability of 71.0$_{-7.7}^{+6.2}$\% 
that the RM of J1243+4043 is greater than the RM of the low-$z$ DLA sample. Such a  probability indicates no significant effect. 
Any additional RM contribution from the plasma of the intragroup medium (IGrM; see Section~\ref{sec:kin}) must be low, 
indicating either low electron density, a weak magnetic field, or field reversals along the line of sight.

The most likely scenarios are either that there is a $(1+z)^2$ dilution in the RM of background quasars, or that the Galactic RM 
foreground varies substantially between samples. There is no evidence for a $(1+z)^2$ dilution in the RM of the DLAs themselves, 
which do not show a marked difference from the control samples. Previous attempts to observe the former scenario of an evolving 
background quasar, e.g. \citet[][]{Hammond12}, find that any $(1+z)^2$ dilution in RM due to cosmological expansion is weak out 
to $z\sim3.5$. If our data do therefore show an evolution in the background quasars, it may be due to a low-redshift selection effect 
on for example SDSS quasars relative to high-redshift SDSS quasars, or possibly a low-redshift evolution in the magnetoionic environment 
surrounding the quasars themselves. 

\begin{figure}
\hbox{
\includegraphics[width=85mm, angle=0]{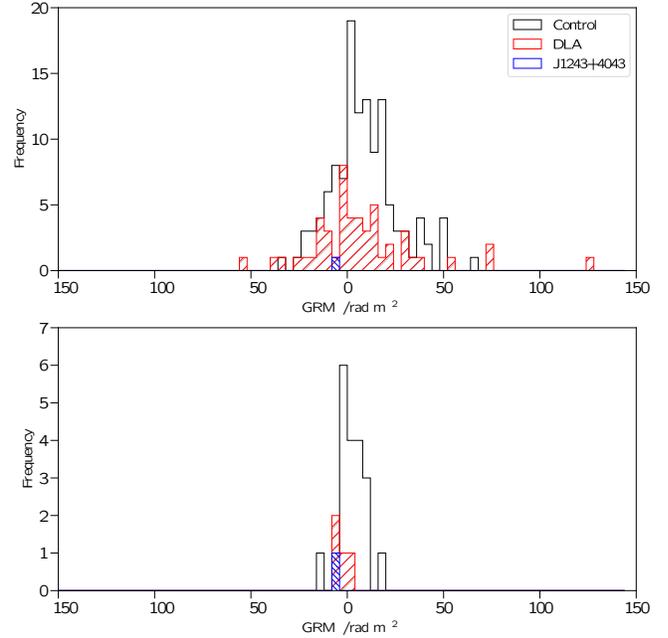}
}
\caption{
The distribution of GRMs for high-$z$ DLAs (top panel) and low-$z$ 21-cm absorbing QGPs (bottom panel).  
The `control' refers to sight lines with no DLA ({\it top}) and no 21-cm absorption ({\it bottom}), respectively.   
}
\label{GRMdist}
\end{figure}
%
To constrain the latter scenario of a varying Galactic RM foreground between samples, we use the best available model from 
\citet[][]{Oppermann15}. A plot of the distribution of Galactic RMs (GRMs) towards each source is shown in Fig.~\ref{GRMdist}. 
There is again no difference between specific samples. We calculate a Bayesian probability of 61.5$_{-17.1}^{+18.5}$\% that the 
GRM towards J1243+4043 is greater than the GRM towards the low-$z$ DLA sample based on 21-cm absorption measurements. This is 
again consistent with no effect. 
The combination of no difference in RM, together with no difference in GRM, means that we are consequently unable to draw 
conclusions about the coherent magnetic fields in the DLA in J1243+4043 itself. The only difference between the datasets are the distributions of the high- 
versus low-redshift RMs, which display no subsequent difference in the GRMs. This is consistent with our earlier interpretation 
of either an uncontrolled selection effect or an evolution of the quasar magnetoionic environment at low redshifts.

In terms of J1243+4043, the magnetic field properties do not allow us to distinguish between different scenarios in order to 
place strong constraints on this source. Future broadband spectropolarimetric observations of this source will enable improved 
tests that can attempt to determine where this sightline fits within the greater source population.

%
\begin{figure*}
\includegraphics[trim=1cm 3.0cm 1.0cm 2.0cm, clip=true, width=175mm, angle=0]{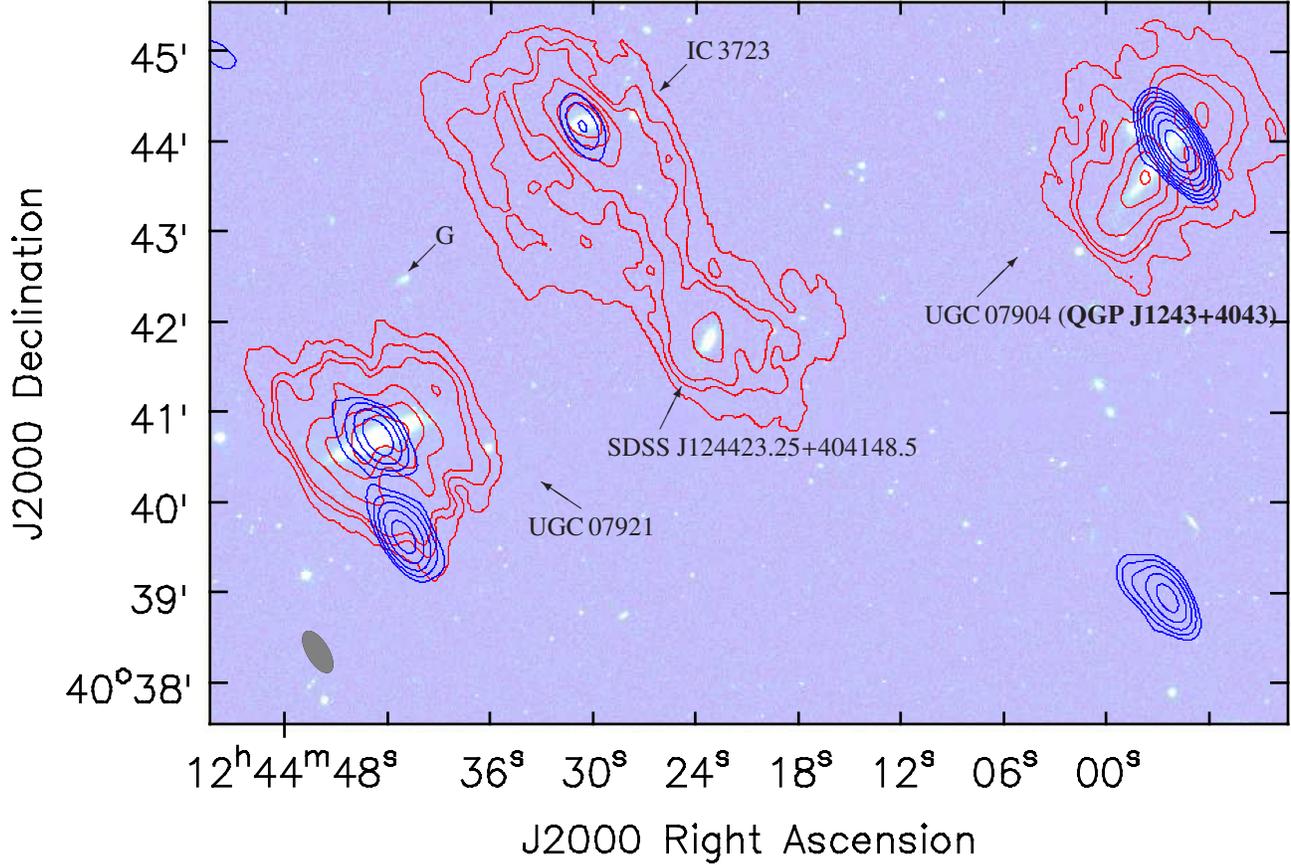}
\caption{
The WSRT radio continuum ({\it blue}) and total \hi-column density contours ({\it red}) overlaid on the 
mosaiced SDSS r-band image.  
The \hi-column density contour levels are 4$\times$10$^{19}$$\times$(1, 1.5, 2, 4, 6, 8, 16)\,cm$^{-2}$.  
The radio continuum contour levels are 1$\times$2$^n$\,mJy\,beam$^{-1}$ (where n=0,1,2,3,..).  
The synthesized beam is 
30.7$^{\prime\prime}$$\times$13.9$^{\prime\prime}$ with P.A. 31$^\circ$.  This corresponds to 
10.5$\times$4.7\,kpc$^2$ at the $z_g$. The QGP J1243+4043 is at {\it top-right} corner.  
The group members are labelled.
}
\vskip -13.5cm
\begin{picture}(400,400)(+10,-50)
\put(170,150){\large UGC\,07921}
\put(190,160){\vector(-3,2){15}}
\put(200,180){\large SDSS J124423.25+404148.5}
\put(220,190){\vector(1,2){08}}
\put(230,330){\large IC\,3723}
\put(230,328){\vector(-1,-1){10}}
\put(135,260){\large G}
\put(135,260){\vector(-1,-1){10}}
\put(320,230){\large UGC\,07904 (\bf {QGP J1243+4043})}
\put(340,240){\vector(1,1){15}}
\end{picture}
\vskip 0.0cm
\label{opthi}
\end{figure*}
%
\subsection[]{\hi\ emission from the associated galaxy group: the environment of the foreground galaxy}
\label{sec:kin}
%
%

\begin{table*}
\centering
\caption{The properties of galaxies in the group}
\begin{tabular}{ccccccccc}
\hline
{\large \strut}  Galaxy name               & Redshift& NUV -r &  $log$\,M$_*$     & log\,SFR     &   $\int$$Sd\nu$  & $log$\,M(\hi) \\    
                                           &         &  (mag) &  (M$_\odot$)  &  (M$_\odot$\,yr$^{-1}$)   &   (Jy\,\kms) & (M$_\odot$)  \\    
\hline                                                                                                                       
{\large \strut}   \ugc\                    & 0.0169  &  1.86  &    9.6   & $-$0.24  & 4.9        & 9.8 \\    
                  SDSS J124423.25+404148.5 & 0.0180  &  2.70  &    9.0   & $-$0.44  &~7.8$^\dag$ & 10$^\dag$  \\    
                  IC\,3723                 & 0.0179  &  1.40  &    9.4   &  ~~0.01  & $-$        &     \\    
                  UGC\,07921 or IC\,3726   & 0.0168  &  3.06  &    10.4  & $-$0.32  & 8.3        & 10  \\    
\hline
\end{tabular}
\flushleft{Note: From left to right, columns show the galaxy name, redshift, NUV - r color, stellar mass, SFR, integrated \hi\ 21-cm line flux and \hi\ mass; 
$\dag$ Total emission line flux density and \hi\ mass of J1244+4041, IC\,3723 and the bridge.}
\label{grphiem}
\end{table*}

In the WSRT data, we serendipitously detect \hi\ 21-cm emission associated with other members of the 
galaxy group.
The group is shown in Fig.~\ref{opthi} which shows the total \hi\ intensity and the radio continuum maps from the WSRT 
data, overlaid on the SDSS r-band image.  We use these images to study the large-scale properties of group members. 
The properties of group members are summarized in Table~\ref{grphiem}.  
For estimating NUV $-$ r colors we have reprocessed GALEX UV photometry following the method described in \citet[][]{Wang10}.  
The colors have been corrected for Galactic extinction following \citet[][]{Wyder07}.
In short, we determine NUV extinction, A$_{\rm {NUV}}$, using A$_{\rm {NUV}}$ - A$_{\rm r}$ = 1.9807\,A$_{\rm r}$, 
where A$_{\rm r}$ is the r-band extinction from SDSS.  
The stellar masses and SFRs have been taken from the Max Planck Institute for Astrophysics/Johns Hopkins University 
value-added catalogues \citep[][]{Kauffmann04, Brinchmann04}. 

The group members are typically low-mass (M$_*$$<$10$^{10}$\,M$_\odot$) blue galaxies with (NUV - r) $<$ 4.
The specific SFR vs M$_*$ trend for the group members is consistent with the \hi-selected sample of 
\citet[][ALFALFA survey]{Huang12}.  
The specific SFR falls off steeply for log\,(M$_*$/M$_\odot$) $>$9.5. The trend in star formation efficiency 
is also consistent with the ALFALFA sample. 

\hi\ 21-cm emission is detected from all four members of the group.  In addition, we detect a $\sim$80\,kpc long \hi\ bridge 
connecting IC\,3723 and SDSS\,J124423.25+404148.5 suggesting that they are interacting.  
The observed, integrated \hi\ emission line flux density of UGC\,07094 $-$ the galaxy associated with the QGP J1243+4043 $-$
is, $\int$$Sdv$ = 4.93\,Jy\,\kms.  The total \hi\ mass, $M_{\rm HI}$ in the units of M$_\odot$, estimated via
\begin{equation}
M_{\rm HI} = \frac{2.356\times10^5}{1+z}\,D_L^2\,\int\,S\,dv,  
\label{eqmhi}
\end{equation}
where $D_L$ is the luminosity distance to the galaxy in Mpc.  
Using this relation and $D_L$=72.7\,Mpc at $z_g$=0.017 we get $M_{\rm HI}$=6$\times$10$^9$\,M$_\odot$. 
The \hi\ mass of UGC\,07921 is $\sim$10$^{10}$\,M$_\odot$. 
The total \hi\ mass of system consisting of SDSS\,J1244+4041, IC\,3723 and the \hi\ bridge 
is 10$^{10}$\,M$_\odot$ (Table~\ref{grphiem}). 

The overall \hi\ content and morphology of a galaxy can be affected by interaction with other galaxies 
\citep[][]{Yun94, Hibbard01, Verdes-Montenegro01, Serra13, Borthakur15, Serra15}. 
In general, the galaxies in high-density environments have less \hi\ than the similar galaxies residing in fields.   
The expected \hi\ masses, log~(M$_{\rm HI}$/M$_\odot$), for UGC\,07904 and UGC\,07921 estimated using 
(NUV - r)-based correlations from \citet[][]{Brown15} are 9.9 and 9.8, respectively.
These correspond to deficiency-parameter, DEF$_{\rm HI}$ = log\,M$_{\rm HI, exp}$ -  log\,M$_{\rm HI, obs}$, of 
only 0.1 and $-$0.2\,dex, respectively.

\begin{figure}
\hbox{
\includegraphics[width=39mm, angle=0]{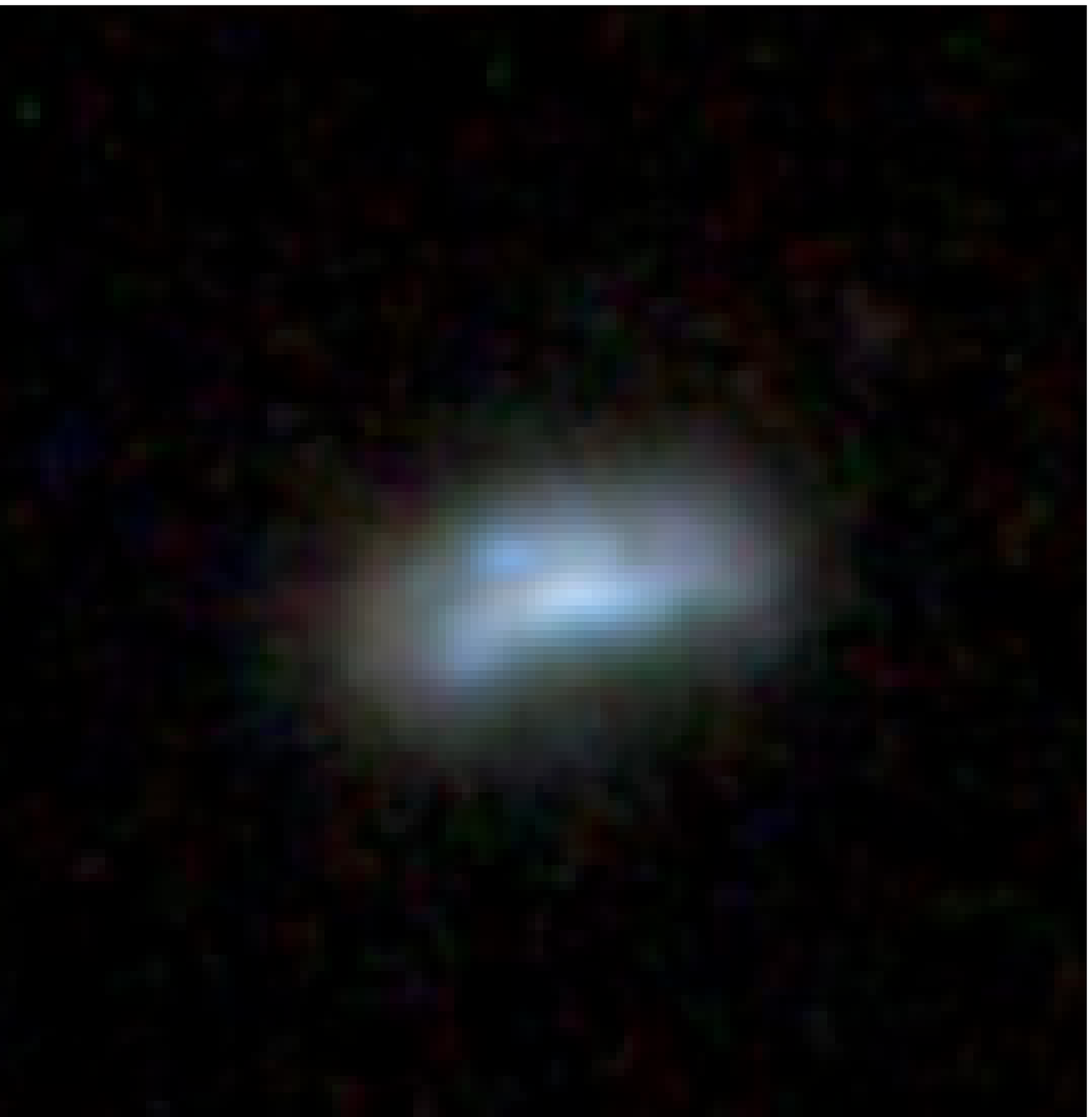}
\includegraphics[width=41mm, angle=0]{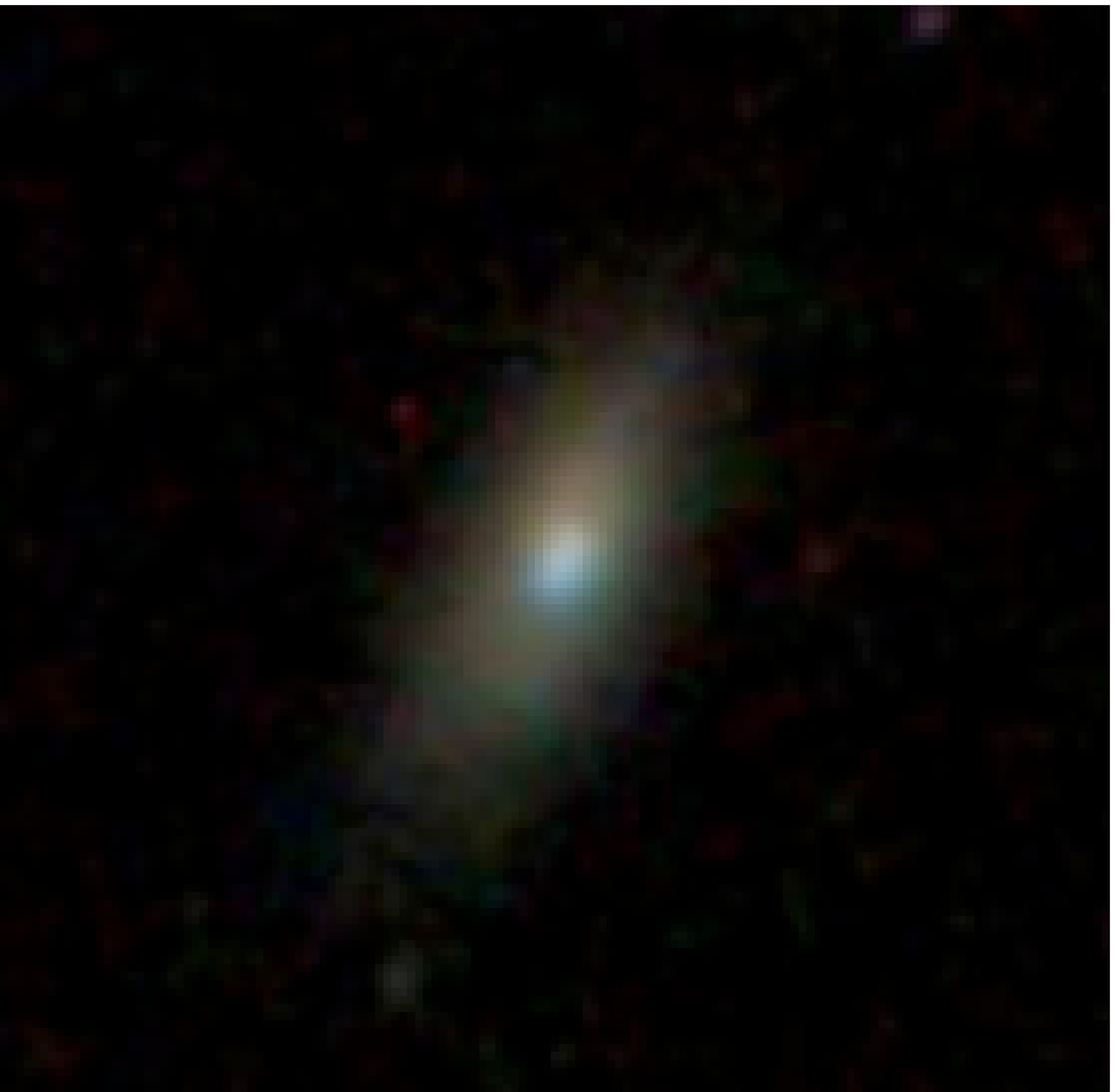}
}
\caption{SDSS color representation (50$^{\prime\prime}$$\times$50$^{\prime\prime}$) 
of IC\,3723 ({\it left}) and J1244+4041.}
\label{sdssgals}
\end{figure}

J1244+4041 and IC\,3723 are embedded in the \hi\ gas stripped by their interaction.  The expected \hi\ mass, 
log~(M$_{\rm HI}$/M$_\odot$), of these two galaxies using the above-mentioned correlations is 9.3 and 9.7, respectively. 
The total observed \hi\ mass of the system consisting of these two galaxies and the \hi\ bridge is remarkably 
consistent with this (Table~\ref{grphiem}). 
Kinematically, both the galaxies have almost identical (within 30\,\kms) recession velocities and we do not observe 
any systematic velocity gradient across the bridge.  Therefore, it is not straightforward to determine their individual \hi\ masses.  
Based on the position-velocity diagrams, in case of IC\,3723 the \hi\ gas is regularly rotating out to a radius of 
20$^{\prime\prime}$ along the major and minor axes.  For J1244+4041, the gas is regularly rotating out to 
35$^{\prime\prime}$ along the major axis and 20$^{\prime\prime}$ along the minor axis.  
Within these apertures the total \hi\ flux associated with J1244+4041 and IC\,3723 is 1.2 and 1.7\,Jy\,\kms, respectively.
The corresponding \hi\ masses, log~(M$_{\rm HI}$/M$_\odot$),  and deficiency parameters, DEF$_{\rm HI}$ , are 9.2 and 9.4, 
and  0.1 and 0.3, respectively.   
Based on the deficiency parameter IC\,3723 has lost most \hi\ gas and is the main contributer to the \hi\ bridge.  
This inference is also supported by its extremely disturbed optical morphology (Fig.~\ref{sdssgals}).  The other group members also have disturbed 
morphologies but to a lesser extent. 
Using SDSS, we identify a faint galaxy (labelled as G in Fig.~\ref{opthi}) at $z$=0.0175 that may also be a member 
of the group.  This galaxy is not detected in \hi\ emission and exhibits distorted morphology.  
More sensitive radio and optical observations, and detailed modelling of SF histories of group members 
is required to determine the complete group membership and understand the details of ongoing interaction.  

In the group, IC\,3723 is bluest and has the highest SFR (1\,M$_\odot$\,yr$^{-1}$), perhaps due to interactions with 
group's members and medium.  Consistent with its enhanced SFR, it is also 
detected in radio continuum with a flux density of 3.8\,mJy.  Following \citet[][]{Bell03}, the SFR inferred from the radio 
continuum is 1.8\,M$_\odot$\,yr$^{-1}$. The SDSS images do not reveal any stellar counterparts to the \hi\ bridge.  
There are two radio continuum sources detected within $\sim$1$^\prime$ of UGC\,07921 (see Fig.~\ref{opthi}).  The one coincident with 
the center of the galaxy has a flux density of 20\,mJy and its southern counterpart has a flux density of 11\,mJy.  Based only 
on the SFR of this galaxy (Table~\ref{grphiem}), we expect to detect a 1.4\,GHz flux density of only $\sim$0.7\,mJy. 
The standard [O\,III]/H$\beta$ and [N\,II]/H$\alpha$ ratios of the galaxy imply that it is a normal star forming 
galaxy \citep[][]{Best12}.  
Therefore, we conclude that the two radio sources are unrelated to UGC\,07921.  It is unclear if these are at a 
higher redshift than the galaxy.  In any case, both the radio sources are too faint and no \hi\ 21-cm absorption is 
detected towards these.  

\section[]{Summary and Conclusions}
\label{sec:summ}

In this paper, we have presented a detailed study of the quasar/galaxy pair: SDSS J124357.5+404346.5 
($z_q$=1.5266)/ UGC\,07904 ($z_g$=0.0169).  
The redshift of the foreground galaxy is from the SDSS whereas the redshift of the quasar is measured using 
the 2-metre optical telescope at IGO. 
The SDSS spectroscopic survey may have missed such interesting quasar sight lines that pass through the 
optical extent of a galaxy \citep[see][for another such case]{Srianand13dib}.  We were able to identify 
these as potential QGPs mainly due to the radio brightness and compactness of AGNs at arcsecond scales.

The measured $\Delta$(g-i) color of the quasar J1243+4043 is -0.05$\pm$0.02.  
Even though the quasar sight line is passing very close to the spiral arm of a galaxy at an impact 
parameter of 6.9\,kpc, the quasar is actually slightly bluer compared to the control sample of 
SDSS quasars at similar redshifts. 
Based on the IGO long-slit spectrum, the metallicity is near-Solar at the center of the 
galaxy, decreases radially outwards, and is about 0.5\,Z$_\odot$ near to the location where 
the quasar sight line passes through the galaxy.  
Overall, the quasar sight line passes through a region of the galaxy with high metallicity and very little   
dust extinction.  

The background quasar, although compact in our GMRT and WSRT observations, it exhibits a core-jet 
morphology with an overall extent of $\sim$9\,pc at $z_g$ in the global VLBI array image. 
We detect \hi\ 21-cm absorption from the foreground galaxy with the GMRT \citep[see also][]{Dutta17}, WSRT, and VLBI data. 
Kinematically, it appears that the 21-cm absorption in this case originates from the gas co-rotating 
with the stellar disk.  
The total \hi\ 21-cm optical depth is 2.24$\pm0.10$. For a spin temperature of 
70\,K  the \hi\ column density is 3$\times$10$^{20}$\,cm$^{-2}$. 
Thus, for temperatures typically seen in the CNM gas in the Galaxy \citep[][]{Heiles03}, the 21-cm absorber
detected towards the quasar can be classified as a DLA.

The VLBI 21-cm absorption spectra are consistent with the GMRT spectrum and show no optical depth variations 
($\tau_{3\sigma} < 0.03$) across the radio source.  
From this we conclude that the size of absorbing clouds are $>$5\,pc and they may be part of diffuse cold gas 
structures that extend beyond $\sim$35\,pc.     
This, together with the cloud sizes inferred from other \hi\ 21-cm absorption measurements from diffuse ISM, suggests 
that the practice of assuming single covering factor to estimate the spin temperature from high-redshift 
DLAs is reasonably justified \citep[e.g.][]{Kanekar09vlba, Gupta12}.  However, caution should be applied in case of 
sight lines tracing the denser phases of the ISM \citep[e.g.][]{Srianand13dib, Biggs16}. Direct measurements of 
clouds sizes using VLBI spectroscopy at high-$z$ with upcoming facilities will be essential for such sight lines
 (see last paragraph). 

Our radio and optical data allow us to constrain the temperature of the absorbing gas using several methods. 
By combining \hi\ column density estimates with the total 21-cm absorption optical depth, 
we determine the harmonic mean spin temperature of the gas to be $\sim$400\,K. 
The 21-cm absorption profile is well-fitted with three Gaussian components.  The width of the narrowest 
component corresponds to a kinetic temperature of 115\,K.  For a simple two-phase medium, adopting this 
as the harmonic mean spin temperature of the CNM, we estimate the CNM-fraction, $f_{CNM}$ = 0.27.  
This is remarkably consistent with the CNM fraction observed in the Galaxy, but less than the high-redshift 
DLAs \citep[][]{Srianand12, Kanekar14}. 

The quasar J1243+4043  is polarized and has a flat spectrum with an integrated 
radio spectral index of $\alpha \sim -$0.3.  Therefore, it offers a unique opportunity to also probe the
magnetoionic plasma from the foreground galaxy along the same line of sight as the 21-cm absorber.  
While the polarized fraction of J1243+4043 is rather typical of the underlying 1.4\,GHz source population, 
the $|{\rm RM}|$  is higher than that of any other source in the high-$z$ DLA sample of \citet[][]{Farnes17}.
A key component of the \citet[][]{Farnes17} data are that the DLAs and background quasars are all located 
at high redshifts, with a median DLA redshift of 2.11 and a quasar redshift of 2.48.
Therefore, we also compare these data with a catalogue of low redshift QGPs with \hi\ 21-cm absorption 
measurements from the literature \citep[][Gupta et al. in prep.]{Dutta17}.  
Critically, we do not find any detectable differences in RMs and polarization fraction of sight lines 
with or without DLAs (or 21-cm absorbers).   
Future broadband spectro-polarimetric observations of the QGP J1243+4043 will enable improved tests that can 
determine where this sight line fits within the greater source population.

The foreground galaxy associated with J1243+4043 is part of a galaxy group.  In our WSRT data, we serendipitously detect 
\hi\ 21-cm emission from four members of the group and 
a $\sim$80\,kpc long \hi\ bridge that connects two of the members.  
Remarkably, the total observed \hi\ mass of these two members and the bridge is consistent with the total 
\hi\ mass of the two galaxies as expected from the (NUV - r) based \hi-scaling relations \citep[][]{Brown15}. 
We find that most of the \hi\ mass to the bridge is contributed by one galaxy.  This particular galaxy perhaps due 
to the interactions with other members of the group, shows bluer colors, higher SFR, and an extremely disturbed 
optical morphology. The other members of the group also have disturbed morphologies but to a lesser extent.

Thanks to large surveys from Square Kilometer Array (SKA) pathfinders and precursors 
\citep[e.g.][]{Allison16spin, Gupta17mals, Maccagni17, Jarvis17} the number of absorption line systems, 
especially sight lines through dusty ISM, at radio wavelengths is expected to dramatically increase over the 
next few years.  Detailed \hi\ 21-cm emission and absorption studies over multiple angular scales as 
presented here will be needed to extract the wealth of information on the neutral ISM in galaxies. 
Such studies will be routinely possible with SKA-VLBI \citep[][]{Paragi15} and the low frequency 
component of ngVLA \citep[][]{Taylor17}.

\section*{Acknowledgments}

We acknowledge useful discussions with Gyula Jozsa and Paolo Serra.
NG acknowledges support from DST Startup Research Grant: {\tt YSS/2014/000338}.   
NG, PN, PP and RS acknowledge support from the Indo-French Centre for the Promotion of Advanced Research 
(Centre Franco-Indien pour la promotion de la recherche avanc\'ee) under Project {\tt 5504-B}.
We acknowledge the use of ARTIP ({\tt https://github.com/RTIP/artip}).  ARTIP is developed by software engineers, 
in particular, Dolly Gyanchandani, Sarang Kulkarni and Ravi Sharma, of ThoughtWorks India Pvt. Limited and 
researchers of IUCAA.  

We thank GMRT, IGO, VLBI (EVN + VLBA) and WSRT staff for their support during the observations.  
GMRT is run by the National Centre for Radio Astrophysics of the Tata Institute of Fundamental Research.
EVN is a joint facility of European, Chinese, South African and other radio astronomy 
institutes funded by their national research councils.
VLBA is run by National Radio Astronomy Observatory.
The National Radio Astronomy Observatory is a facility of the National Science Foundation 
operated under cooperative agreement by Associated Universities, Inc. 
The research leading to these results has received funding from the European Commission 7th
Framework Programme (FP/2007-2013) under grant agreement No. 283393 (RadioNet3).
Scientific results from the VLBI data presented in this publication are derived from the following 
EVN project code: {\tt GG074}.  
WSRT is operated by the ASTRON (Netherlands Institute for Radio Astronomy) with 
support from the Netherlands Foundation for Scientific Research (NWO).
We acknowledge the use of SDSS spectra from the archive (http://www.sdss.org/).
Funding for the SDSS and SDSS-II has been provided by the Alfred P. Sloan Foundation, 
the Participating Institutions, the National Science Foundation, the U.S. Department of Energy, 
the National Aeronautics and Space Administration, the Japanese Monbukagakusho, the Max Planck Society, 
and the Higher Education Funding Council for England. 
The Image Reduction and Analysis Facility (IRAF) is distributed by the National 
Optical Astronomy Observatory, which is operated by the Association of Universities 
for Research in Astronomy (AURA) under a cooperative agreement with the National Science 
Foundation.
The Common Astronomy Software Applications (CASA) package is 
developed by an international consortium of scientists based at the National Radio 
Astronomical Observatory (NRAO), the European Southern Observatory (ESO), the National 
Astronomical Observatory of Japan (NAOJ), the Academia Sinica Institute of Astronomy 
and Astrophysics (ASIAA), the CSIRO division for Astronomy and Space Science (CASS), 
and the Netherlands Institute for Radio Astronomy (ASTRON) under the guidance of NRAO. 
The Astronomical Image Processing System (AIPS) 
is produced and maintained by the National Radio Astronomy Observatory, a facility of the 
National Science Foundation operated under cooperative agreement by Associated Universities, Inc.

\def\aj{AJ}%
\def\actaa{Acta Astron.}%
\def\araa{ARA\&A}%
\def\apj{ApJ}%
\def\apjl{ApJ}%
\def\apjs{ApJS}%
\def\ao{Appl.~Opt.}%
\def\apss{Ap\&SS}%
\def\aap{A\&A}%
\def\aapr{A\&A~Rev.}%
\def\aaps{A\&AS}%
\def\azh{AZh}%
\def\baas{BAAS}%
\def\bac{Bull. astr. Inst. Czechosl.}%
\def\caa{Chinese Astron. Astrophys.}%
\def\cjaa{Chinese J. Astron. Astrophys.}%
\def\icarus{Icarus}%
\def\jcap{J. Cosmology Astropart. Phys.}%
\def\jrasc{JRASC}%
\def\mnras{MNRAS}%
\def\memras{MmRAS}%
\def\na{New A}%
\def\nar{New A Rev.}%
\def\pasa{PASA}%
\def\pra{Phys.~Rev.~A}%
\def\prb{Phys.~Rev.~B}%
\def\prc{Phys.~Rev.~C}%
\def\prd{Phys.~Rev.~D}%
\def\pre{Phys.~Rev.~E}%
\def\prl{Phys.~Rev.~Lett.}%
\def\pasp{PASP}%
\def\pasj{PASJ}%
\def\qjras{QJRAS}%
\def\rmxaa{Rev. Mexicana Astron. Astrofis.}%
\def\skytel{S\&T}%
\def\solphys{Sol.~Phys.}%
\def\sovast{Soviet~Ast.}%
\def\ssr{Space~Sci.~Rev.}%
\def\zap{ZAp}%
\def\nat{Nature}%
\def\iaucirc{IAU~Circ.}%
\def\aplett{Astrophys.~Lett.}%
\def\apspr{Astrophys.~Space~Phys.~Res.}%
\def\bain{Bull.~Astron.~Inst.~Netherlands}%
\def\fcp{Fund.~Cosmic~Phys.}%
\def\gca{Geochim.~Cosmochim.~Acta}%
\def\grl{Geophys.~Res.~Lett.}%
\def\jcp{J.~Chem.~Phys.}%
\def\jgr{J.~Geophys.~Res.}%
\def\jqsrt{J.~Quant.~Spec.~Radiat.~Transf.}%
\def\memsai{Mem.~Soc.~Astron.~Italiana}%
\def\nphysa{Nucl.~Phys.~A}%
\def\physrep{Phys.~Rep.}%
\def\physscr{Phys.~Scr}%
\def\planss{Planet.~Space~Sci.}%
\def\procspie{Proc.~SPIE}%
\let\astap=\aap
\let\apjlett=\apjl
\let\apjsupp=\apjs
\let\applopt=\ao
\bibliographystyle{mn}
\bibliography{/Users/ngupta/Desktop/Comp/mybib}

\begin{thebibliography}{80}
\expandafter\ifx\csname natexlab\endcsname\relax\def\natexlab#1{#1}\fi

\bibitem[{{Allison} {et~al.}(2017){Allison}, {Moss}, {Macquart}, {Curran},
  {Duchesne}, {Mahony}, {Sadler}, {Whiting}, {Bannister}, {Chippendale},
  {Edwards}, {Harvey-Smith}, {Heywood}, {Indermuehle}, {Lenc}, {Marvil},
  {McConnell}, \& {Sault}}]{Allison16}
{Allison}, J.~R., {Moss}, V.~A., {Macquart}, J.-P., {et~al.}, 2017, \mnras,
  465, 4450

\bibitem[{{Allison} {et~al.}(2016){Allison}, {Zwaan}, {Duchesne}, \&
  {Curran}}]{Allison16spin}
{Allison}, J.~R., {Zwaan}, M.~A., {Duchesne}, S.~W., \& {Curran}, S.~J., 2016,
  \mnras, 462, 1341

\bibitem[{{Argence} \& {Lamareille}(2009)}]{Argence09}
{Argence}, B. \& {Lamareille}, F., 2009, \aap, 495, 759

\bibitem[{{Bell}(2003)}]{Bell03}
{Bell}, E.~F., 2003, \apj, 586, 794

\bibitem[{{Bernet} {et~al.}(2008){Bernet}, {Miniati}, {Lilly}, {Kronberg}, \&
  {Dessauges-Zavadsky}}]{Bernet08}
{Bernet}, M.~L., {Miniati}, F., {Lilly}, S.~J., {Kronberg}, P.~P., \&
  {Dessauges-Zavadsky}, M., 2008, \nat, 454, 302

\bibitem[{{Best} \& {Heckman}(2012)}]{Best12}
{Best}, P.~N. \& {Heckman}, T.~M., 2012, \mnras, 421, 1569

\bibitem[{{Biggs} {et~al.}(2016){Biggs}, {Zwaan}, {Hatziminaoglou},
  {P{\'e}roux}, \& {Liske}}]{Biggs16}
{Biggs}, A.~D., {Zwaan}, M.~A., {Hatziminaoglou}, E., {P{\'e}roux}, C., \&
  {Liske}, J., 2016, \mnras, 462, 2819

\bibitem[{{Boiss\'e} {et~al.}(1998){Boiss\'e}, {Le Brun}, {Bergeron}, \&
  {Deharveng}}]{Boisse98}
{Boiss\'e}, P., {Le Brun}, V., {Bergeron}, J., \& {Deharveng}, J.-M., 1998,
  \aap, 333, 841

\bibitem[{{Borthakur}(2016)}]{Borthakur16}
{Borthakur}, S., 2016, \apj, 829, 128

\bibitem[{{Borthakur} {et~al.}(2011){Borthakur}, {Tripp}, {Yun}, {Bowen},
  {Meiring}, {York}, \& {Momjian}}]{Borthakur11}
{Borthakur}, S., {Tripp}, T.~M., {Yun}, M.~S., {Bowen}, D.~V., {Meiring},
  J.~D., {York}, D.~G., \& {Momjian}, E., 2011, \apj, 727, 52

\bibitem[{{Borthakur} {et~al.}(2010){Borthakur}, {Tripp}, {Yun}, {Momjian},
  {Meiring}, {Bowen}, \& {York}}]{Borthakur10}
{Borthakur}, S., {Tripp}, T.~M., {Yun}, M.~S., {Momjian}, E., {Meiring}, J.~D.,
  {Bowen}, D.~V., \& {York}, D.~G., 2010, \apj, 713, 131

\bibitem[{{Borthakur} {et~al.}(2015){Borthakur}, {Yun}, {Verdes-Montenegro},
  {Heckman}, {Zhu}, \& {Braatz}}]{Borthakur15}
{Borthakur}, S., {Yun}, M.~S., {Verdes-Montenegro}, L., {Heckman}, T.~M.,
  {Zhu}, G., \& {Braatz}, J.~A., 2015, \apj, 812, 78

\bibitem[{{Braun}(2012)}]{Braun12}
{Braun}, R., 2012, \apj, 749, 87

\bibitem[{{Briggs} \& {Wolfe}(1983)}]{Briggs83}
{Briggs}, F.~H. \& {Wolfe}, A.~M., 1983, \apj, 268, 76

\bibitem[{{Brinchmann} {et~al.}(2004){Brinchmann}, {Charlot}, {White},
  {Tremonti}, {Kauffmann}, {Heckman}, \& {Brinkmann}}]{Brinchmann04}
{Brinchmann}, J., {Charlot}, S., {White}, S.~D.~M., {Tremonti}, C.,
  {Kauffmann}, G., {Heckman}, T., \& {Brinkmann}, J., 2004, \mnras, 351, 1151

\bibitem[{{Brown} {et~al.}(2015){Brown}, {Catinella}, {Cortese}, {Kilborn},
  {Haynes}, \& {Giovanelli}}]{Brown15}
{Brown}, T., {Catinella}, B., {Cortese}, L., {Kilborn}, V., {Haynes}, M.~P., \&
  {Giovanelli}, R., 2015, \mnras, 452, 2479

\bibitem[{{Carilli} {et~al.}(1996){Carilli}, {Lane}, {de Bruyn}, {Braun}, \&
  {Miley}}]{Carilli96}
{Carilli}, C.~L., {Lane}, W., {de Bruyn}, A.~G., {Braun}, R., \& {Miley},
  G.~K., 1996, \aj, 111, 1830

\bibitem[{{Carilli} \& {van Gorkom}(1992)}]{Carilli92}
{Carilli}, C.~L. \& {van Gorkom}, J.~H., 1992, \apj, 399, 373

\bibitem[{{Curran}(2017)}]{Curran17}
{Curran}, S.~J., 2017, \mnras, 470, 3159

\bibitem[{{Curran} {et~al.}(2013){Curran}, {Allison}, {Glowacki}, {Whiting}, \&
  {Sadler}}]{Curran13}
{Curran}, S.~J., {Allison}, J.~R., {Glowacki}, M., {Whiting}, M.~T., \&
  {Sadler}, E.~M., 2013, \mnras, 431, 3408

\bibitem[{{Curran} {et~al.}(2010){Curran}, {Tzanavaris}, {Darling}, {Whiting},
  {Webb}, {Bignell}, {Athreya}, \& {Murphy}}]{Curran10}
{Curran}, S.~J., {Tzanavaris}, P., {Darling}, J.~K., {Whiting}, M.~T., {Webb},
  J.~K., {Bignell}, C., {Athreya}, R., \& {Murphy}, M.~T., 2010, \mnras, 402,
  35

\bibitem[{{Dutta} {et~al.}(2016){Dutta}, {Gupta}, {Srianand}, \&
  {O'Meara}}]{Dutta16}
{Dutta}, R., {Gupta}, N., {Srianand}, R., \& {O'Meara}, J.~M., 2016, \mnras,
  456, 4209

\bibitem[{{Dutta} {et~al.}(2017{\natexlab{a}}){Dutta}, {Srianand}, {Gupta}, \&
  {Joshi}}]{Dutta17mg2}
{Dutta}, R., {Srianand}, R., {Gupta}, N., \& {Joshi}, R., 2017{\natexlab{a}},
  \mnras, 468, 1029

\bibitem[{{Dutta} {et~al.}(2017{\natexlab{b}}){Dutta}, {Srianand}, {Gupta},
  {Momjian}, {Noterdaeme}, {Petitjean}, \& {Rahmani}}]{Dutta17}
{Dutta}, R., {Srianand}, R., {Gupta}, N., {Momjian}, E., {Noterdaeme}, P.,
  {Petitjean}, P., \& {Rahmani}, H., 2017{\natexlab{b}}, \mnras, 465, 588

\bibitem[{{Farnes} {et~al.}(2014){Farnes}, {O'Sullivan}, {Corrigan}, \&
  {Gaensler}}]{Farnes14}
{Farnes}, J.~S., {O'Sullivan}, S.~P., {Corrigan}, M.~E., \& {Gaensler}, B.~M.,
  2014, \apj, 795, 63

\bibitem[{{Farnes} {et~al.}(2017){Farnes}, {Rudnick}, {Gaensler}, {Haverkorn},
  {O'Sullivan}, \& {Curran}}]{Farnes17}
{Farnes}, J.~S., {Rudnick}, L., {Gaensler}, B.~M., {Haverkorn}, M.,
  {O'Sullivan}, S.~P., \& {Curran}, S.~J., 2017, \apj, 841, 67

\bibitem[{{Gupta} {et~al.}(2017){Gupta}, {Srianand}, {Baan}, {Baker},
  {Beswick}, {Bhatnagar}, {Bhattacharya}, {Bosma}, {Carilli}, {Cluver},
  {Combes}, {Cress}, {Dutta}, {Fynbo}, {Heald}, {Hilton}, {Hussain}, {Jarvis},
  {Jozsa}, {Kamphuis}, {Kembhavi}, {Kerp}, {Kl{\"o}ckner}, {Krogager},
  {Kulkarni}, {Ledoux}, {Mahabal}, {Mauch}, {Moodley}, {Momjian}, {Morganti},
  {Noterdaeme}, {Oosterloo}, {Petitjean}, {Schr{\"o}der}, {Serra}, {Sievers},
  {Spekkens}, {V{\"a}is{\"a}nen}, {van der Hulst}, {Vivek}, {Wang}, {Wong}, \&
  {Zungu}}]{Gupta17mals}
{Gupta}, N., {Srianand}, R., {Baan}, W., {et~al.}, 2017, ArXiv e-prints

\bibitem[{{Gupta} {et~al.}(2010){Gupta}, {Srianand}, {Bowen}, {York}, \&
  {Wadadekar}}]{Gupta10}
{Gupta}, N., {Srianand}, R., {Bowen}, D.~V., {York}, D.~G., \& {Wadadekar}, Y.,
  2010, \mnras, 408, 849

\bibitem[{{Gupta} {et~al.}(2013){Gupta}, {Srianand}, {Noterdaeme}, {Petitjean},
  \& {Muzahid}}]{Gupta13}
{Gupta}, N., {Srianand}, R., {Noterdaeme}, P., {Petitjean}, P., \& {Muzahid},
  S., 2013, \aap, 558, A84

\bibitem[{{Gupta} {et~al.}(2012){Gupta}, {Srianand}, {Petitjean}, {Bergeron},
  {Noterdaeme}, \& {Muzahid}}]{Gupta12}
{Gupta}, N., {Srianand}, R., {Petitjean}, P., {Bergeron}, J., {Noterdaeme}, P.,
  \& {Muzahid}, S., 2012, \aap, 544, A21

\bibitem[{{Gupta} {et~al.}(2009){Gupta}, {Srianand}, {Petitjean}, {Noterdaeme},
  \& {Saikia}}]{Gupta09}
{Gupta}, N., {Srianand}, R., {Petitjean}, P., {Noterdaeme}, P., \& {Saikia},
  D.~J., 2009, \mnras, 398, 201

\bibitem[{{Hammond} {et~al.}(2012){Hammond}, {Robishaw}, \&
  {Gaensler}}]{Hammond12}
{Hammond}, A.~M., {Robishaw}, T., \& {Gaensler}, B.~M., 2012, ArXiv e-prints

\bibitem[{{Haschick} {et~al.}(1983){Haschick}, {Crane}, \& {Baan}}]{Haschick83}
{Haschick}, A.~D., {Crane}, P.~C., \& {Baan}, W.~A., 1983, \apjl, 269, L43

\bibitem[{{Heiles} \& {Troland}(2003)}]{Heiles03}
{Heiles}, C. \& {Troland}, T.~H., 2003, \apj, 586, 1067

\bibitem[{{Helmboldt} {et~al.}(2007){Helmboldt}, {Taylor}, {Tremblay},
  {Fassnacht}, {Walker}, {Myers}, {Sjouwerman}, {Pearson}, {Readhead},
  {Weintraub}, {Gehrels}, {Romani}, {Healey}, {Michelson}, {Blandford}, \&
  {Cotter}}]{Helmboldt07}
{Helmboldt}, J.~F., {Taylor}, G.~B., {Tremblay}, S., {et~al.}, 2007, \apj, 658,
  203

\bibitem[{{Hibbard} {et~al.}(2001){Hibbard}, {van der Hulst}, {Barnes}, \&
  {Rich}}]{Hibbard01}
{Hibbard}, J.~E., {van der Hulst}, J.~M., {Barnes}, J.~E., \& {Rich}, R.~M.,
  2001, \aj, 122, 2969

\bibitem[{{Hoppmann} {et~al.}(2015){Hoppmann}, {Staveley-Smith}, {Freudling},
  {Zwaan}, {Minchin}, \& {Calabretta}}]{Hoppmann15}
{Hoppmann}, L., {Staveley-Smith}, L., {Freudling}, W., {Zwaan}, M.~A.,
  {Minchin}, R.~F., \& {Calabretta}, M.~R., 2015, \mnras, 452, 3726

\bibitem[{{Huang} {et~al.}(2012){Huang}, {Haynes}, {Giovanelli}, \&
  {Brinchmann}}]{Huang12}
{Huang}, S., {Haynes}, M.~P., {Giovanelli}, R., \& {Brinchmann}, J., 2012,
  \apj, 756, 113

\bibitem[{{Jarvis} {et~al.}(2017){Jarvis}, {Taylor}, {Agudo}, {Allison},
  {Deane}, {Frank}, {Gupta}, {Heywood}, {Maddox}, {McAlpine}, {Santos},
  {Scaife}, {Vaccari}, {Zwart}, {Adams}, {Bacon}, {Baker}, {Bassett}, {Best},
  {Beswick}, {Blyth}, {Brown}, {Bruggen}, {Cluver}, {Colafranceso}, {Cotter},
  {Cress}, {Dave}, {Ferrari}, {Hardcastle}, {Hale}, {Harrison}, {Hatfield},
  {Klockner}, {Kolwa}, {Malefahlo}, {Marubini}, {Mauch}, {Moodley}, {Morganti},
  {Norris}, {Peters}, {Prandoni}, {Prescott}, {Oliver}, {Oozeer}, {Rottgering},
  {Seymour}, {Simpson}, {Smirnov}, {Smith}, {Spekkens}, {Stil}, {Tasse}, {van
  der Heyden}, {Whittam}, \& {WIlliams}}]{Jarvis17}
{Jarvis}, M.~J., {Taylor}, A.~R., {Agudo}, I., {et~al.}, 2017, ArXiv e-prints

\bibitem[{{Kanekar} {et~al.}(2009{\natexlab{a}}){Kanekar}, {Lane}, {Momjian},
  {Briggs}, \& {Chengalur}}]{Kanekar09vlba}
{Kanekar}, N., {Lane}, W.~M., {Momjian}, E., {Briggs}, F.~H., \& {Chengalur},
  J.~N., 2009{\natexlab{a}}, \mnras, 394, L61

\bibitem[{{Kanekar} {et~al.}(2009{\natexlab{b}}){Kanekar}, {Prochaska},
  {Ellison}, \& {Chengalur}}]{Kanekar09mg2}
{Kanekar}, N., {Prochaska}, J.~X., {Ellison}, S.~L., \& {Chengalur}, J.~N.,
  2009{\natexlab{b}}, \mnras, 396, 385

\bibitem[{{Kanekar} {et~al.}(2014){Kanekar}, {Prochaska}, {Smette}, {Ellison},
  {Ryan-Weber}, {Momjian}, {Briggs}, {Lane}, {Chengalur}, {Delafosse}, {Grave},
  {Jacobsen}, \& {de Bruyn}}]{Kanekar14}
{Kanekar}, N., {Prochaska}, J.~X., {Smette}, A., {et~al.}, 2014, \mnras, 438,
  2131

\bibitem[{{Kauffmann} {et~al.}(2004){Kauffmann}, {White}, {Heckman},
  {M{\'e}nard}, {Brinchmann}, {Charlot}, {Tremonti}, \&
  {Brinkmann}}]{Kauffmann04}
{Kauffmann}, G., {White}, S.~D.~M., {Heckman}, T.~M., {M{\'e}nard}, B.,
  {Brinchmann}, J., {Charlot}, S., {Tremonti}, C., \& {Brinkmann}, J., 2004,
  \mnras, 353, 713

\bibitem[{{Keeney} {et~al.}(2005){Keeney}, {Momjian}, {Stocke}, {Carilli}, \&
  {Tumlinson}}]{Keeney05}
{Keeney}, B.~A., {Momjian}, E., {Stocke}, J.~T., {Carilli}, C.~L., \&
  {Tumlinson}, J., 2005, \apj, 622, 267

\bibitem[{{Kennicutt}(1998{\natexlab{a}})}]{Kennicutt98}
{Kennicutt}, Jr., R.~C., 1998{\natexlab{a}}, \araa, 36, 189

\bibitem[{{Kennicutt}(1998{\natexlab{b}})}]{Kennicutt98b}
---, 1998{\natexlab{b}}, \apj, 498, 541

\bibitem[{{Kronberg} \& {Perry}(1982)}]{Kronberg82}
{Kronberg}, P.~P. \& {Perry}, J.~J., 1982, \apj, 263, 518

\bibitem[{{Lane} {et~al.}(2000){Lane}, {Briggs}, \& {Smette}}]{Lane00}
{Lane}, W.~M., {Briggs}, F.~H., \& {Smette}, A., 2000, \apj, 532, 146

\bibitem[{{Liszt}(2001)}]{Liszt01}
{Liszt}, H., 2001, \aap, 371, 698

\bibitem[{{Mac Low} \& {Klessen}(2004)}]{Maclow04}
{Mac Low}, M.-M. \& {Klessen}, R.~S., 2004, Reviews of Modern Physics, 76, 125

\bibitem[{{Maccagni} {et~al.}(2017){Maccagni}, {Morganti}, {Oosterloo},
  {Ger{\'e}b}, \& {Maddox}}]{Maccagni17}
{Maccagni}, F.~M., {Morganti}, R., {Oosterloo}, T.~A., {Ger{\'e}b}, K., \&
  {Maddox}, N., 2017, \aap, 604, A43

\bibitem[{{Madau} \& {Dickinson}(2014)}]{Madau14}
{Madau}, P. \& {Dickinson}, M., 2014, \araa, 52, 415

\bibitem[{{Malik} {et~al.}(2017){Malik}, {Chand}, \& {Seshadri}}]{Malik17}
{Malik}, S., {Chand}, H., \& {Seshadri}, T.~R., 2017, ArXiv e-prints

\bibitem[{{Nilson}(1973)}]{Nilson73}
{Nilson}, P., 1973, Nova Acta Regiae Soc.~Sci.~Upsaliensis Ser.~V, 0

\bibitem[{{Noterdaeme} {et~al.}(2012){Noterdaeme}, {Petitjean}, {Carithers},
  {P{\^a}ris}, {Font-Ribera}, {Bailey}, {Aubourg}, {Bizyaev}, {Ebelke},
  {Finley}, {Ge}, {Malanushenko}, {Malanushenko}, {Miralda-Escud{\'e}},
  {Myers}, {Oravetz}, {Pan}, {Pieri}, {Ross}, {Schneider}, {Simmons}, \&
  {York}}]{Noterdaeme12dla}
{Noterdaeme}, P., {Petitjean}, P., {Carithers}, W.~C., {et~al.}, 2012, \aap,
  547, L1

\bibitem[{{Oppermann} {et~al.}(2015){Oppermann}, {Junklewitz}, {Greiner},
  {En{\ss}lin}, {Akahori}, {Carretti}, {Gaensler}, {Goobar}, {Harvey-Smith},
  {Johnston-Hollitt}, {Pratley}, {Schnitzeler}, {Stil}, \&
  {Vacca}}]{Oppermann15}
{Oppermann}, N., {Junklewitz}, H., {Greiner}, M., {et~al.}, 2015, \aap, 575,
  A118

\bibitem[{{Osterbrock} \& {Ferland}(2006)}]{Osterbrock06}
{Osterbrock}, D.~E. \& {Ferland}, G.~J., 2006, {Astrophysics of gaseous nebulae
  and active galactic nuclei}

\bibitem[{{Paragi} {et~al.}(2015){Paragi}, {Godfrey}, {Reynolds}, {Rioja},
  {Deller}, {Zhang}, {Gurvits}, {Bietenholz}, {Szomoru}, {Bignall}, {Boven},
  {Charlot}, {Dodson}, {Frey}, {Garrett}, {Imai}, {Lobanov}, {Reid}, {Ros},
  {van Langevelde}, {Zensus}, {Zheng}, {Alberdi}, {Agudo}, {An}, {Argo},
  {Beswick}, {Biggs}, {Brunthaler}, {Campbell}, {Cimo}, {Colomer}, {Corbel},
  {Conway}, {Cseh}, {Deane}, {Falcke}, {Gawronski}, {Gaylard}, {Giovannini},
  {Giroletti}, {Goddi}, {Goedhart}, {G{\'o}mez}, {Gunn}, {Kharb}, {Kloeckner},
  {Koerding}, {Kovalev}, {Kunert-Bajraszewska}, {Lindqvist}, {Lister},
  {Mantovani}, {Marti-Vidal}, {Mezcua}, {McKean}, {Middelberg}, {Miller-Jones},
  {Moldon}, {Muxlow}, {O'Brien}, {Perez-Torres}, {Pogrebenko}, {Quick},
  {Rushton}, {Schilizzi}, {Smirnov}, {Sohn}, {Surcis}, {Taylor}, {Tingay},
  {Tudose}, {van der Horst}, {van Leeuwen}, {Venturi}, {Vermeulen},
  {Vlemmings}, {de Witt}, {Wucknitz}, {Yang}, {Gab{\"a}nyi}, \&
  {Jung}}]{Paragi15}
{Paragi}, Z., {Godfrey}, L., {Reynolds}, C., {et~al.}, 2015, Advancing
  Astrophysics with the Square Kilometre Array (AASKA14), 143

\bibitem[{{Pettini} \& {Pagel}(2004)}]{Pettini04a}
{Pettini}, M. \& {Pagel}, B.~E.~J., 2004, \mnras, 348, L59

\bibitem[{{Reeves} {et~al.}(2015){Reeves}, {Sadler}, {Allison}, {Koribalski},
  {Curran}, \& {Pracy}}]{Reeves15}
{Reeves}, S.~N., {Sadler}, E.~M., {Allison}, J.~R., {Koribalski}, B.~S.,
  {Curran}, S.~J., \& {Pracy}, M.~B., 2015, \mnras, 450, 926

\bibitem[{{Reeves} {et~al.}(2016){Reeves}, {Sadler}, {Allison}, {Koribalski},
  {Curran}, {Pracy}, {Phillips}, {Bignall}, \& {Reynolds}}]{Reeves16}
{Reeves}, S.~N., {Sadler}, E.~M., {Allison}, J.~R., {et~al.}, 2016, \mnras,
  457, 2613

\bibitem[{{Richards} {et~al.}(2003){Richards}, {Hall}, {Vanden Berk},
  {Strauss}, {Schneider}, {Weinstein}, {Reichard}, {York}, {Knapp}, {Fan},
  {Ivezi{\'c}}, {Brinkmann}, {Budav{\'a}ri}, {Csabai}, \&
  {Nichol}}]{Richards03}
{Richards}, G.~T., {Hall}, P.~B., {Vanden Berk}, D.~E., {et~al.}, 2003, \aj,
  126, 1131

\bibitem[{{Serra} {et~al.}(2013){Serra}, {Koribalski}, {Duc}, {Oosterloo},
  {McDermid}, {Michel-Dansac}, {Emsellem}, {Cuillandre}, {Alatalo}, {Blitz},
  {Bois}, {Bournaud}, {Bureau}, {Cappellari}, {Crocker}, {Davies}, {Davis}, {de
  Zeeuw}, {Khochfar}, {Krajnovi{\'c}}, {Kuntschner}, {Lablanche}, {Morganti},
  {Naab}, {Sarzi}, {Scott}, {Weijmans}, \& {Young}}]{Serra13}
{Serra}, P., {Koribalski}, B., {Duc}, P.-A., {et~al.}, 2013, \mnras, 428, 370

\bibitem[{{Serra} {et~al.}(2015){Serra}, {Koribalski}, {Kilborn}, {Allison},
  {Amy}, {Ball}, {Bannister}, {Bell}, {Bock}, {Bolton}, {Bowen}, {Boyle},
  {Broadhurst}, {Brodrick}, {Brothers}, {Bunton}, {Chapman}, {Cheng},
  {Chippendale}, {Chung}, {Cooray}, {Cornwell}, {DeBoer}, {Diamond}, {Forsyth},
  {Gough}, {Gupta}, {Hampson}, {Harvey-Smith}, {Hay}, {Hayman}, {Heywood},
  {Hotan}, {Hoyle}, {Humphreys}, {Indermuehle}, {Jacka}, {Jackson}, {Jackson},
  {Jeganathan}, {Johnston}, {Joseph}, {Kamphuis}, {Leach}, {Lenc}, {Lensson},
  {Mackay}, {Marquarding}, {Marvil}, {McClure-Griffiths}, {McConnell}, {Meyer},
  {Mirtschin}, {Neuhold}, {Ng}, {Norris}, {O'Sullivan}, {Pathikulangara},
  {Pearce}, {Phillips}, {Popping}, {Qiao}, {Reynolds}, {Roberts}, {Sault},
  {Schinckel}, {Shaw}, {Shimwell}, {Staveley-Smith}, {Storey}, {Sweetnam},
  {Troup}, {Tzioumis}, {Voronkov}, {Westmeier}, {Whiting}, {Wilson}, {Wong}, \&
  {Wu}}]{Serra15}
{Serra}, P., {Koribalski}, B., {Kilborn}, V., {et~al.}, 2015, \mnras, 452, 2680

\bibitem[{{Srianand} {et~al.}(2012{\natexlab{a}}){Srianand}, {Gupta},
  {Petitjean}, {Noterdaeme}, {Ledoux}, {Salter}, \& {Saikia}}]{Srianand12dla}
{Srianand}, R., {Gupta}, N., {Petitjean}, P., {Noterdaeme}, P., {Ledoux}, C.,
  {Salter}, C.~J., \& {Saikia}, D.~J., 2012{\natexlab{a}}, \mnras, 421, 651

\bibitem[{{Srianand} {et~al.}(2012{\natexlab{b}}){Srianand}, {Gupta},
  {Petitjean}, {Noterdaeme}, {Ledoux}, {Salter}, \& {Saikia}}]{Srianand12}
---, 2012{\natexlab{b}}, \mnras, 421, 651

\bibitem[{{Srianand} {et~al.}(2013){Srianand}, {Gupta}, {Rahmani}, {Momjian},
  {Petitjean}, \& {Noterdaeme}}]{Srianand13dib}
{Srianand}, R., {Gupta}, N., {Rahmani}, H., {Momjian}, E., {Petitjean}, P., \&
  {Noterdaeme}, P., 2013, \mnras, 428, 2198

\bibitem[{{Taylor} {et~al.}(2009){Taylor}, {Stil}, \& {Sunstrum}}]{Taylor09}
{Taylor}, A.~R., {Stil}, J.~M., \& {Sunstrum}, C., 2009, \apj, 702, 1230

\bibitem[{{Taylor} {et~al.}(2017){Taylor}, {Dowell}, {Malins}, {Clarke},
  {Kassim}, {Giacintucci}, {Hicks}, {Kooi}, {Peters}, {Polisensky}, {Schinzel},
  \& {Stovall}}]{Taylor17}
{Taylor}, G., {Dowell}, J., {Malins}, J., {et~al.}, 2017, ArXiv e-prints

\bibitem[{{Vanden Berk} {et~al.}(2001){Vanden Berk}, {Richards}, {Bauer},
  {Strauss}, {Schneider}, {Heckman}, {York}, {Hall}, {Fan}, {Knapp},
  {Anderson}, {Annis}, {Bahcall}, {Bernardi}, {Briggs}, {Brinkmann}, {Brunner},
  {Burles}, {Carey}, {Castander}, {Connolly}, {Crocker}, {Csabai}, {Doi},
  {Finkbeiner}, {Friedman}, {Frieman}, {Fukugita}, {Gunn}, {Hennessy},
  {Ivezi{\'c}}, {Kent}, {Kunszt}, {Lamb}, {Leger}, {Long}, {Loveday}, {Lupton},
  {Meiksin}, {Merelli}, {Munn}, {Newberg}, {Newcomb}, {Nichol}, {Owen}, {Pier},
  {Pope}, {Rockosi}, {Schlegel}, {Siegmund}, {Smee}, {Snir}, {Stoughton},
  {Stubbs}, {SubbaRao}, {Szalay}, {Szokoly}, {Tremonti}, {Uomoto}, {Waddell},
  {Yanny}, \& {Zheng}}]{VandenBerk01}
{Vanden Berk}, D.~E., {Richards}, G.~T., {Bauer}, A., {et~al.}, 2001, \aj, 122,
  549

\bibitem[{{Verdes-Montenegro} {et~al.}(2001){Verdes-Montenegro}, {Yun},
  {Williams}, {Huchtmeier}, {Del Olmo}, \& {Perea}}]{Verdes-Montenegro01}
{Verdes-Montenegro}, L., {Yun}, M.~S., {Williams}, B.~A., {Huchtmeier}, W.~K.,
  {Del Olmo}, A., \& {Perea}, J., 2001, \aap, 377, 812

\bibitem[{{Vivek} {et~al.}(2009){Vivek}, {Srianand}, {Noterdaeme}, {Mohan}, \&
  {Kuriakosde}}]{Vivek09}
{Vivek}, M., {Srianand}, R., {Noterdaeme}, P., {Mohan}, V., \& {Kuriakosde},
  V.~C., 2009, \mnras, 400, L6

\bibitem[{{Wang} {et~al.}(2010){Wang}, {Overzier}, {Kauffmann}, {von der
  Linden}, \& {Kong}}]{Wang10}
{Wang}, J., {Overzier}, R., {Kauffmann}, G., {von der Linden}, A., \& {Kong},
  X., 2010, \mnras, 401, 433

\bibitem[{{White} {et~al.}(1999){White}, {Bliton}, {Bhavsar}, {Bornmann},
  {Burns}, {Ledlow}, \& {Loken}}]{White99}
{White}, R.~A., {Bliton}, M., {Bhavsar}, S.~P., {Bornmann}, P., {Burns}, J.~O.,
  {Ledlow}, M.~J., \& {Loken}, C., 1999, \aj, 118, 2014

\bibitem[{{Wild} {et~al.}(2007){Wild}, {Kauffmann}, {Heckman}, {Charlot},
  {Lemson}, {Brinchmann}, {Reichard}, \& {Pasquali}}]{Wild07a}
{Wild}, V., {Kauffmann}, G., {Heckman}, T., {Charlot}, S., {Lemson}, G.,
  {Brinchmann}, J., {Reichard}, T., \& {Pasquali}, A., 2007, \mnras, 381, 543

\bibitem[{{Wolfe} {et~al.}(1992){Wolfe}, {Lanzetta}, \& {Oren}}]{Wolfe92}
{Wolfe}, A.~M., {Lanzetta}, K.~M., \& {Oren}, A.~L., 1992, \apj, 388, 17

\bibitem[{{Wolfire} {et~al.}(1995){Wolfire}, {Hollenbach}, {McKee}, {Tielens},
  \& {Bakes}}]{Wolfire95}
{Wolfire}, M.~G., {Hollenbach}, D., {McKee}, C.~F., {Tielens}, A.~G.~G.~M., \&
  {Bakes}, E.~L.~O., 1995, \apj, 443, 152

\bibitem[{{Wyder} {et~al.}(2007){Wyder}, {Martin}, {Schiminovich}, {Seibert},
  {Budav{\'a}ri}, {Treyer}, {Barlow}, {Forster}, {Friedman}, {Morrissey},
  {Neff}, {Small}, {Bianchi}, {Donas}, {Heckman}, {Lee}, {Madore}, {Milliard},
  {Rich}, {Szalay}, {Welsh}, \& {Yi}}]{Wyder07}
{Wyder}, T.~K., {Martin}, D.~C., {Schiminovich}, D., {et~al.}, 2007, \apjs,
  173, 293

\bibitem[{{York} {et~al.}(2006){York}, {Khare}, {Vanden Berk}, {Kulkarni},
  {Crotts}, {Lauroesch}, {Richards}, {Schneider}, {Welty}, {Alsayyad}, {Kumar},
  {Lundgren}, {Shanidze}, {Smith}, {Vanlandingham}, {Baugher}, {Hall},
  {Jenkins}, {Menard}, {Rao}, {Tumlinson}, {Turnshek}, {Yip}, \&
  {Brinkmann}}]{York06}
{York}, D.~G., {Khare}, P., {Vanden Berk}, D., {et~al.}, 2006, \mnras, 367, 945

\bibitem[{{Yun} {et~al.}(1994){Yun}, {Ho}, \& {Lo}}]{Yun94}
{Yun}, M.~S., {Ho}, P.~T.~P., \& {Lo}, K.~Y., 1994, \nat, 372, 530

\end{thebibliography}


\end{document}